\documentclass[journal]{IEEEtran}
\IEEEoverridecommandlockouts
\usepackage{color}

\usepackage{enumerate}
\usepackage[shortlabels]{enumitem}

\usepackage[pdftex]{graphicx}

\usepackage{caption}
\usepackage{subcaption}

\usepackage{tikz}
\usetikzlibrary{calc, arrows, shapes, positioning, intersections, decorations.markings, decorations.pathmorphing, matrix}
\usepackage{tikz-cd}

\usepackage{siunitx}

\usepackage[plain]{algorithm}
\usepackage{algpseudocode}

\usepackage{pgfplots}
\pgfplotsset{compat=1.3}

\usepackage{dsfont}

\usepackage{cite}

\usepackage{multicol}

\usepackage{bm}

\usepackage{flushend}

\usepackage{acronym}
\acrodef{phd}[PHD]{Probability Hypothesis Density}
\acrodef{cphd}[CPHD]{Cardinalized PHD}
\acrodef{fov}[FoV]{Field of View}
\acrodef{nb}[NB]{Negative Binomial}
\acrodef{gm}[GM]{Gaussian Mixture}
\acrodef{pgfl}[PGFL]{Probability Generating Functional}
\acrodef{iid}[i.i.d.]{independent and identically distributed}
\acrodef{wrt}[w.r.t.]{with respect to}
\acrodef{fisst}[FISST]{Finite Set Statistics}
\acrodef{epsrc}[EPSRC]{Engineering and Physical Sciences Research Council}
\acrodef{esric}[ESRIC]{Edinburgh Super-Resolution Imaging Consortium}
\acrodef{smc}[SMC]{Sequential Monte Carlo}
\acrodef{rfs}[RFS]{Random Finite Set}
\acrodef{mc}[MC]{Monte Carlo}
\acrodef{ospa}[OSPA]{Optimal Sub-Pattern Assignment}
\acrodef{mht}[MHT]{Multi-Hypothesis Tracking}
\acrodef{jpda}[JPDA]{Joint Probabilistic Data Association}
\acrodef{member}[MeMBer]{Multi-Bernoulli}

\usepackage[cmex10]{amsmath}
\makeatletter
\newcommand{\pushright}[1]{\ifmeasuring@#1\else\omit\hfill$\displaystyle#1$\fi\ignorespaces}
\newcommand{\pushleft}[1]{\ifmeasuring@#1\else\omit$\displaystyle#1$\hfill\fi\ignorespaces}

\makeatother

\usepackage{mathtools}
\usepackage{amssymb}
\usepackage{amsthm}
\newtheorem{thm}{Theorem}[section]
\newtheorem{assum}[thm]{Assumptions}
\newtheorem{lem}[thm]{Lemma}
\newtheorem{cor}[thm]{Corollary}
\newtheorem{prop}[thm]{Proposition}

\newcommand{\sumneq}{\sideset{}{^{\neq}}\sum}

\usepackage{multirow, multicol}
\usepackage{fixltx2e}

\def\b{\mathrm{b}}
\def\Bcal{\mathcal{B}}
\def\c{\mathrm{c}}
\def\corr{\mathrm{corr}}
\def\cov{\mathrm{cov}}
\def\d{\mathrm{d}}
\def\Exp{\mathbb{E}}
\def\Fcal{\mathcal{F}}
\def\Gcal{\mathcal{G}}
\def\J{\mathrm{J}}
\def\Lcal{\mathcal{L}}
\def\Nbb{\mathbb{N}}
\def\Ncal{\mathcal{N}}

\def\Pbb{\mathbb{P}}
\def\Rbb{\mathbb{R}}
\def\s{\mathrm{s}}
\def\var{\mathrm{var}}

\def\Xcal{\mathcal{X}}
\def\Xfk{\mathfrak{X}}
\def\Zcal{\mathcal{Z}}
  
\title{A second-order PHD filter with mean and variance in target number} 

\author{\IEEEauthorblockN{Isabel Schlangen\IEEEauthorrefmark{1},
Emmanuel Delande\IEEEauthorrefmark{1},
J\'er\'emie Houssineau\IEEEauthorrefmark{1} and
Daniel E. Clark\IEEEauthorrefmark{1}}
\thanks{$^*$ School of Electrical and Physical Sciences,
Heriot-Watt University, Edinburgh EH14 4AS, UK. Email: \{is117, E.D.Delande, J.Houssineau, D.E.Clark\}@hw.ac.uk. Isabel Schlangen is supported by the Edinburgh Super-Resolution Imaging Consortium (MR/K01563X/1). This work was supported by the EPSRC Platform Grant (EP/J015180/1), and the MoD UDRC Phase 2 (EP/K014227/1). J. Houssineau was sponsored by the French DCNS.}}

\begin{document}

\maketitle

\begin{abstract}
The \ac{phd} and \ac{cphd} filters are popular solutions to the multi-target tracking problem due to their low complexity and ability to estimate the number and states of targets in cluttered environments. The \ac{phd} filter propagates the first-order moment (i.e. mean) of the number of targets while the \ac{cphd} propagates the cardinality distribution in the number of targets, albeit for a greater computational cost. Introducing the Panjer point process, this paper proposes a second-order  \ac{phd} filter, propagating the second-order moment (i.e. variance) of the number of targets alongside its mean. The resulting algorithm is more versatile in the modelling choices than the \ac{phd} filter, and its computational cost is significantly lower compared to the  \ac{cphd} filter. The paper compares the three filters in statistical simulations which demonstrate that the proposed filter reacts more quickly to changes in the number of targets, i.e., target births and target deaths, than the \ac{cphd} filter. In addition, a new statistic for multi-object filters is introduced in order to study the correlation between the estimated number of targets in different regions of the state space, and propose a quantitative analysis of the spooky effect for the three filters.
\end{abstract}

\IEEEpeerreviewmaketitle

\acresetall
\section{Introduction}

In the context of multi-target detection and tracking problems, methods based on the \ac{rfs} framework have recently attracted a lot of attention due to the development of low-complexity algorithms within this methodology \cite{Mahler2007statistical}. The best-known algorithm is perhaps the \ac{phd} filter that jointly estimates the number of targets and their states by propagating the first-order moment of a \ac{rfs} \cite{Mahler2003Multitarget}; a \ac{gm} and a \ac{smc} implementation have been presented in \cite{Vo2006Gaussian} and \cite{Vo2005Sequential}.

Erdinc and Willett \cite{Erdinc2005Probability} suggested that only propagating the first-order moment did not provide sufficient information for applications where a high confidence in the target number was needed. Consequently, Mahler derived the \ac{cphd} filter which propagates the cardinality distribution of the target point process alongside its first-order moment\cite{Mahler2007PHD}. It thus provides higher-order information on the number of targets, but to the expense of a higher computational cost. Around the same time, he also proposed a filter restricted to the first two moments using a binomial approximation \cite{Mahler2006PHD}. However, due to the binomial approximation it was suggested that restrictions were required on the relative number of false alarms and targets.  In 2007, Vo et al.~showed that the \ac{cphd} filter can be overconfident in some cases \cite{Vo2007Analytic}, and in 2009, Fr{\"a}nken et al.~identified a counter-intuitive property of the \ac{cphd} filter that occurs with the weights of the targets when they are miss-detected which they called the spooky effect \cite{Franken2009Spooky}. An alternative approach for extending the \ac{phd} filter to a second-order filter was proposed by Singh et al.~using a Gauss-Poisson prior  \cite{Singh2009Filters}. 

Other developments in the \ac{fisst} framework have focussed on more advanced filtering solutions. The \ac{member} filter \cite{Mahler2007statistical} is based on a fully Bayesian approach where the system assumes that each target is modelled by a state estimate and a probability of existence. The bias in the number of targets in the original \ac{member} filter was addressed in \cite{Vo2009cardinality}, and further developments around Bernoulli \acp{rfs} were introduced in \cite{Williams2012Hybrid,Williams2015efficient}. Various methods propagating information on individual targets within the \ac{fisst} framework have been developed since \cite{Vo2013Labeled,Vo2014Labeled,Reuter2014labeled}.

In this paper, we introduce a second-order \ac{phd} filter in which the predicted target process is assumed Panjer instead of Poisson. The Panjer distribution \cite{Fackler2009Panjer} is specified by two parameters and encompasses the binomial, Poisson and negative binomial distributions; unlike a Poisson distribution, it can describe a population of targets whose estimated size has a higher or lower variance than its mean. The proposed solution thus complements the original \ac{phd} filter with the variance in the estimated number of targets; it also propagates less information than the \ac{cphd} filter but has a lower computational cost. The Panjer distribution was studied for the analysis of the \ac{cphd} update in \cite{Franken2009Spooky}, though it was not used to develop a new filter. The proposed filter can also be seen as a generalisation of the \ac{phd} filter with a negative binomial-distributed false alarms \cite{Schlangen2016PHD}, which was designed for tracking scenarios with high variability in background noise. We also exploit the statistical tools introduced in this paper in order to study the correlation in the estimated number of targets in disjoint regions of the state space, and provide a quantitative analysis of the well-known spooky effect \cite{Franken2009Spooky} for the \ac{phd} filter, \ac{cphd} filter, and the proposed second-order \ac{phd} filter.

Sec.~\ref{sec:background} introduces some background material on point processes. The description of four relevant point processes follows in Sec.~\ref{sec:pointprocessesexamples}, then used in Sec.~\ref{sec:filter} to formulate the recursion of the proposed second-order \ac{phd} filter. The construction of the regional correlation for the \ac{phd}, second-order \ac{phd}, and \ac{cphd} filters is detailed in Sec.~\ref{sec:correlation}. A comparison of the \ac{gm} implementations of the three filters is then given in Sec.~\ref{sec:experiments}, and Sec.~\ref{sec:conclusion} concludes. Pseudo-code and detailed proofs for the proposed algorithms are given in the appendix.

\section{Background}
\label{sec:background}
Many recent works in multi-object filtering exploit Mahler's \ac{fisst} framework \cite{Mahler2007statistical}, in which multi-target state configurations are described by \acp{rfs}. The \ac{fisst} framework allows for the production of the densities of various statistical quantities describing a \ac{rfs} (multi-object density, \acl{phd}, etc.) through the set derivative operator.

This paper considers higher-order statistical quantities whose expression arises naturally from probability \emph{measures} rather than \emph{densities}, such as the regional covariance, or does not admit a density altogether, such as the regional variance or correlation (see Sec. \ref{subsec:moment}). Hence we shall favour the measure-theoretical formulation originating from the point process theory, for which a specific methodology has been developed to construct higher-order statistical moment measures or densities through the chain derivative operator \cite{Bernhard2005Chain}.

In the section, we provide the necessary background material on point processes, and highlight the connections with the \ac{fisst} framework when appropriate. For the rest of the paper, $(\Omega, \Fcal, \Pbb)$ denotes a probability space with sample space $\Omega$, $\sigma$-algebra $\Fcal$, and probability measure $\Pbb$. Throughout the paper, all random variables are defined on $(\Omega,\Fcal,\Pbb)$ and we denote by $\Exp$ the expectation \ac{wrt} $\Pbb$.

\subsection{Point processes}
\label{subsec:pointprocesses}
We denote by $\Xcal \subseteq \Rbb^{d_x}$ the $d_x$-dimensional state space describing the state of an individual object (position, velocity, etc.). A point process $\Phi$ on $\Xcal$ is a random variable on the process space $\Xfk = \bigcup_{n\geq 0}\Xcal^n$, i.e., the space of finite sequences of points in $\Xcal$. A realisation of $\Phi$ is a sequence ${\varphi = (x_1,\dots,x_n)\in \Xcal^n}$, representing a population of $n$ objects with states $x_i \in \Xcal$. Point processes can be described using their probability distribution $P_\Phi$ on the measurable space $(\Xfk,\Bcal(\Xfk))$, where $\Bcal(\Xfk)$ denotes the Borel $\sigma$-algebra of the process space $\Xfk$ \cite{Stoyan1997Stochastic}. 

The projection measure $P^{(n)}_\Phi$ of the probability distribution $P_\Phi$ on $\Xcal^n$, $n \geq 0$, describes the realisations of $\Phi$ with $n$ elements; the projection measures of a point process are always defined as symmetrical functions, so that the permutations of a realisation $\varphi$ are equally probable. Furthermore, a point process is called \emph{simple} if $\varphi$ does not contain repetitions, i.e.~its elements are pairwise distinct almost surely. For the rest of the paper, all point processes are assumed simple. In that case, it is assumed that the probability distribution $P_\Phi$ of a point process admits a density $p_\Phi$ \ac{wrt} some reference measure $\lambda$. The densities of the projection measures $P_{\Phi}^{(n)}$ are denoted by $p_{\Phi}^{(n)}$, and both quantities will be exploited throughout the paper.

In the literature originating from Mahler's \ac{fisst} framework \cite{Mahler2003Multitarget, Vo2006Gaussian}, an alternative construction of simple point processes is a \ac{rfs}, a random object whose realizations are \emph{sets} of points $\{x_1,\ldots, x_n\}$, in which the elements are by construction \emph{unordered}.

\subsection{Multi-target Bayesian filtering} \label{subsec:bayes}
In the context of multi-target tracking, we make use of a \emph{target point process} $\Phi_k$ to describe the information about the target population at time $k$. The scene is observed by a sensor system, providing sets of measurements at discrete times (indexed by $k\in \mathbb{N}$ in the following). The $d_z$-dimensional observation space describing the individual measurements produced by the sensor (range, azimuth, etc.) is denoted by $\Zcal\subseteq \mathbb{R}^{d_z}$. The set of measurements collected at time $k$ is denoted by $Z_k$.

Point processes can be cast into a Bayesian framework in order to propagate $\Phi_k$ over time \cite{Mahler2007statistical}. Bayesian filtering consists of a \emph{prediction} or \emph{time update} step which is concerned with the motion model, birth and death of targets, and a \emph{data update} step which models the observation process, missed detections and false alarms and exploits the current measurement set $Z_k$.

The full multi-target Bayesian recursion propagates the law $P_k$ of the target process $\Phi_k$. The time prediction and data update equations at time $k$ are given by \cite{Mahler2007statistical}
      \begin{align}
	P_{k|k-1}(\d\xi) &= \int T_{k|k-1}(\d\xi|\varphi) P_{k-1}(\d\varphi), \label{EqMultiBayesFilter1}
	\\
	P_{k}(\d\xi|Z_k) &= \frac{L_k(Z_k|\xi)P_{k|k-1}(\d\xi)}{\int L_k(Z_k|\varphi) P_{k|k-1}(\d\varphi)}, \label{EqMultiBayesFilter2}
      \end{align}
where $T_{k|k-1}$ is the multi-target Markov transition kernel from time $k - 1$ to time $k$, and $L_k$ is the multi-measurement/multi-target likelihood at time step $k$.\footnote{When $\mu$, $\mu'$ are two measures on some space $X$, we use the notation $\mu(\d x) = \mu'(\d x)$, where $x\in X$, to indicate that $\int f(x)\mu(\d x) = \int f(x)\mu'(\d x)$ for any bounded measurable function $f$ on $X$.} Note that the formulation of the multi-target Bayesian recursion with \emph{measure-theoretical} integrals \eqref{EqMultiBayesFilter1}, \eqref{EqMultiBayesFilter2} is drawn from its original \ac{rfs}-based formulation in \cite{Mahler2007statistical} with \emph{set} integrals.

\subsection{Statistical moments}
\label{subsec:moment}
Similarly to real-valued random variables, statistical moments can be defined for a point process $\Phi$ in order to provide an alternative description to its probability distribution $P_{\Phi}$ (or, equivalently, to its projection measures $P^{(n)}_\Phi$ for any $n \in \Nbb$). Statistical moments will play an important role in this paper, for the construction of the second-order \ac{phd} filter in Sec.~\ref{sec:filter} as well as for the study of the correlation in the estimated target number in distinct regions of the state space in Sec.~\ref{sec:correlation}.

The \emph{$n$-th order moment measure} $\mu_\Phi^{(n)}$ of a point process $\Phi$ is the measure on $\Xcal^n$ such that, for any bounded measurable function $f_n$ on $\Xcal^n$, it holds that \cite{Stoyan1997Stochastic}
\begin{equation}
	\int f_n(x_{1:n}) \mu^{(n)}_{\Phi}(\d(x_{1:n})) = \Exp\bigg[\sum_{x_1,\dots,x_n \in \Phi}f_n(x_{1:n})\bigg] \label{eq:nonfacmeasure}
\end{equation}
where we use the shorter notation $x_{1:n}$ to denote the sequence $(x_1,\dots,x_n)$.\footnote{When $\varphi \in X^n$, $n \geq 0$, is a sequence of elements on some space $X$, the abuse of notation ``$x \in \varphi$'' is used to denote that the element $x \in X$ appears in the sequence $\varphi$.} In addition, the \emph{$n$-th order factorial moment measure} $\nu_\Phi^{(n)}$ of a point process $\Phi$ is the measure on $\Xcal^n$ such that, for any bounded measurable function $f_n$ on $\Xcal^n$, it holds that \cite{Stoyan1997Stochastic}
\begin{equation}
    \int f_n(x_{1:n}) \nu^{(n)}_{\Phi}(\d(x_{1:n})) = \Exp\bigg[~~\sumneq_{x_1,\dots,x_n \in \Phi}f_n(x_{1:n})\bigg] \label{eq:facmeasure}
\end{equation}
where $\Sigma^{\neq}$ indicates that the selected points $x_1,\dots,x_n$ are all pairwise distinct. It can be shown that for any bounded measurable function $f_n$ on $\Xcal^n$, it holds that
\begin{equation}
 \int \left[\sum_{x_1,\ldots,x_n \in \varphi} f_n(x_{1:n})\right] P_{\Phi}(\d\varphi) = \int f_n(x_{1:n}) \nu^{(n)}_{\Phi}(\d x_{1:n}). \label{eq:campbell}
\end{equation}
This result is known as Campbell's theorem \cite{Stoyan1997Stochastic}.

Setting $f_n(x_{1:n}) = \prod_{i=1}^n \mathds{1}_{B_i}(x_i)$ in Eqs~\eqref{eq:nonfacmeasure}, \eqref{eq:facmeasure}, yields
\begin{align}
\mu^{(n)}_\Phi(B_1 \!\times \dots \times\! B_n) &= \mathbb{E}\bigg[ \hspace{-0.6cm}\sum_{~~~~~x_1,\dots,x_n \in \Phi}\hspace{-0.7cm}\mathds{1}_{B_1}(x_1) \dots  \mathds{1}_{B_n}(x_n)\bigg],\label{eq:nmomentmeasure}
\\
\nu^{(n)}_\Phi(B_1\! \times \dots \times\! B_n) &= \mathbb{E}\bigg[ {\hspace{-0.4cm}\sumneq_{~~~~~x_1,\dots,x_n \in \Phi}}\hspace{-0.6cm}\mathds{1}_{B_1}(x_1) \dots \mathds{1}_{B_n}(x_n)\bigg],\label{eq:nmomentfactorialmeasure}
\end{align}
for any regions $B_i \in \Bcal(\Xcal)$, $1 \leq i \leq n$.\footnote{The notation $\mathds{1}_B$ denotes the indicator function, i.e., $\mathds{1}_B(x) = 1 $ if $x \in B$, and zero otherwise.}
Eqs~\eqref{eq:nmomentmeasure} and \eqref{eq:nmomentfactorialmeasure} provide some insight on the moment measures. The scalar $\mu^{(n)}_\Phi(B_1 \!\times \dots \times\! B_n)$  estimates the joint localisation of sequence points within the regions $B_i$, while $\nu^{(n)}_{\Phi}(B_1 \!\times \dots \times\! B_n)$ further imposes the sequence points to be pairwise distinct. 

Note that the first-order moment measure $\mu_{\Phi}^{(1)}$ coincides with the first-order factorial moment measure $\nu_{\Phi}^{(1)}$; it is known as the \emph{intensity measure} of the point process and simply denoted by $\mu_{\Phi}$. Its associated density, also denoted by $\mu_{\Phi}$, is called the \emph{intensity} of the point process $\Phi$, more usually called \emph{Probability Hypothesis Density} in the context of \acp{rfs} \cite{Mahler2003Multitarget}. In this paper we shall also exploit the second-order moment measures; similarly to real-valued random variables we can define the \emph{covariance}, \emph{variance}, and \emph{correlation} of a point process $\Phi$ as \cite{Stoyan1997Stochastic,Illian2008Statistical}
\begin{align}
  \cov_{\Phi}(B, B') &= \mu_{\Phi}^{(2)}(B\times B') - \mu_{\Phi}(B)\mu_{\Phi}(B'), \label{eq:covariance}
  \\
  \var_{\Phi}(B) &= \mu_{\Phi}^{(2)}(B\times B) - \left[\mu_{\Phi}(B)\right]^2, \label{eq:variance}
  \\
  \corr_{\Phi}(B, B') &= \cfrac{\cov_\Phi(B,B')}{\sqrt{\var_\Phi(B)}\sqrt{\var_\Phi(B')}}, \label{eq:correlation}
\end{align}
for any regions $B,B'\in \Bcal(\Xcal)$. The scalar $\mu_{\Phi}(B)$ yields the expected (or \emph{mean}) number of objects within $B$, while the scalar $\var_{\Phi}(B)$ quantifies the spread of the estimated number of objects within $B$ around its mean value $\mu_{\Phi}(B)$ \cite{Delande2014Regional}. Finally, the scalar $\corr_{\Phi}(B, B')$ quantifies the correlation between the estimated number of targets within $B$ and $B'$; it will be exploited in this paper to assess the so-called ``spooky effect'' of multi-object filters, coined in \cite{Franken2009Spooky} for the \ac{cphd} filter.

Note that in the general case the variance $\var_{\Phi}$ is a non-additive function, and does not admit a density. Note also that the second-order moment measure can be decomposed into the sum
\begin{equation}
\label{eq:decomp}
\mu_\Phi^{(2)}(B\times B') = \mu_\Phi(B\cap B') + \nu_\Phi^{(2)}(B \times B'),
\end{equation}
for any regions $B, B' \in \Bcal(\Xcal)$.

\subsection{Point processes and functionals}
\label{subsec:functionals}
Similarly to the Fourier transform for signals or the probability generating function for discrete real-valued random variables, convenient tools exist to handle operations on point processes. 
The \emph{Laplace functional} $\Lcal_\Phi$ and the \emph{\ac{pgfl}} $\Gcal_\Phi$ of a point process $\Phi$ are defined by
\begin{align}
\Lcal_\Phi(f) &= \sum_{n\geq 0} \int \exp\left(-\sum_{i=1}^n f(x_i)\right)P^{(n)}_\Phi(\d x_{1:n}), \label{eq:laplace}
\\
\Gcal_\Phi(h) &= \sum_{n\geq 0} \int \left[ \prod_{i=1}^n h(x_i) \right]P^{(n)}_\Phi(\d x_{1:n}), \label{eq:pgfl}
\end{align}
respectively for two test functions $f:\Xcal \rightarrow \mathbb{R}^+$ and $h: \Xcal \rightarrow [0,1]$. Note that from \eqref{eq:laplace} and \eqref{eq:pgfl} it holds that
\begin{equation}
\Gcal_\Phi(h) = \Lcal_\Phi(-\ln h).
\end{equation}
Depending on the nature of the point process $\Phi$, the expression of the functionals may reduce to simpler expressions that do not involve infinite sums (see examples in Sec.~\ref{sec:pointprocessesexamples}).

\subsection{Point processes and differentiation}
\label{subsec:functionaldiff}
In this paper we shall exploit the \emph{chain differential} \cite{Bernhard2005Chain},  a convenient operator that allows for the evaluation of both the statistical moments of a point process $\Phi$ and their corresponding densities through the differentiation of its Laplace functional $\Lcal_{\Phi}$ or its \ac{pgfl} $\Gcal_{\Phi}$ \cite{Clark2012Generalized,Clark2013Faa,Clark2015few}.

Given a functional $G$ and two functions $h,\eta: \Xcal \rightarrow \mathbb{R}^+$, the (chain) differential of $G$ \ac{wrt} $h$ in the direction of $\eta$ is defined as \cite{Bernhard2005Chain}
\begin{equation}
\label{eq:chaindiff}
\delta G(h;\eta) = \lim_{n\rightarrow\infty}\cfrac{G(h+\varepsilon_n \eta_n) - G(h)}{\varepsilon_n},
\end{equation}
when the limit exists and is identical for any sequence of real numbers $(\varepsilon_n)_{n\in \mathbb{N}}$ converging to $0$ and any sequence of functions $(\eta_n: \Xcal \rightarrow \mathbb{R}^+)_{n\in \mathbb{N}}$ converging pointwise to $\eta$. 

The statistical quantities described in Sec.~\ref{subsec:pointprocesses} and Sec.~\ref{subsec:moment} can then be extracted through the following differentiations:
\begin{align}
P_\Phi^{(n)}(B_1\!\times \cdots \times\!B_n) &= \frac{1}{n!}
\delta^n\Gcal_\Phi(h;\mathds{1}_{B_1}, \ldots, \mathds{1}_{B_n})|_{h=0}, \label{eq:probaderivative}
\\
\mu_\Phi^{(n)}(B_1\!\times \cdots \times\!B_n) &= (-1)^n \delta^n\Lcal_\Phi(f; \mathds{1}_{B_1}, \ldots, \mathds{1}_{B_n})|_{f=0}, \label{eq:muderivative}
\\
\nu_\Phi^{(n)}(B_1\!\times \cdots \times\!B_n) &= \delta^n\Gcal_\Phi(h;\mathds{1}_{B_1}, \ldots, \mathds{1}_{B_n})|_{h=1}, \label{eq:nuderivative}
\end{align}
for any regions $B_i \in \Bcal(\Xcal)$, $1 \leq i \leq n$ \cite{Stoyan1997Stochastic}. The chain differential has convenient properties and leads to similar rules to the classical derivative: namely, a \emph{product rule}  \cite{Bernhard2005Chain}
\begin{equation}
\label{eq:productrule}
\delta(F\cdot G)(h;\eta) = \delta F(h;\eta) G(h) + F(h) \delta G(h;\eta),
\end{equation}
and a \emph{chain rule} \cite{Bernhard2005Chain}
\begin{equation}
\label{eq:chainrule}
\delta(F\circ G)(h;\eta) = \delta F(G(h);\delta G(h;\eta)).
\end{equation}
They can be generalised to the \emph{$n$-fold product rule} \cite{Clark2015few}
\begin{equation}
\label{eq:generalproductrule}
\begin{split}
&\delta^n(F\cdot G)(h;\eta_1,\dots,\eta_n) \\
&= \sum_{\omega\subseteq \{1,\dots,n\}} \delta^{|\omega|} F\Big(h;(\eta_i)_{i \in \omega}\Big) \delta^{|\omega^{\c}|} G\Big(h;(\eta_j)_{j \in {\omega}^\c}\Big),
\end{split}
\end{equation}
where ${\omega}^\c = \{1,\dots,n\}\setminus \omega$ is the complement of $\omega$, and the \emph{$n$-fold chain rule} or \emph{Fa\`a di Bruno's formula for chain differentials}~\cite{Clark2013Faa, Clark2015few}
\begin{equation}
\label{eq:generalchainrule}
\begin{split}
\delta^n&(F\circ G)(h;\eta_1,\dots,\eta_n) \\
&= \sum_{\pi \in \Pi_n} \delta^{|\pi|} F\left(G(h);\left(\delta^{|\omega|}G(h; (\eta_i)_{i \in \omega})\right)_{\omega \in \pi}\right),
\end{split}
\end{equation}
where $\Pi_n$ is the set of partitions of the index set $\{1,\dots,n\}$. The equivalent of the $n$-fold product rule \eqref{eq:generalproductrule} in the \ac{fisst} framework is called the generalised product rule for set derivatives \cite[p.~389]{Mahler2007statistical}. Fa\`a di Bruno's formula \eqref{eq:generalchainrule} has recently been applied for spatial cluster modelling~\cite{Clark2013Faa2}, Volterra series~\cite{clark2014Faa}, multi-target spawning~\cite{Bryant2016CPHD}, and for negative binomial clutter modelling~\cite{Schlangen2016PHD}.

When the chain differential \eqref{eq:chaindiff} is linear and continuous \ac{wrt} its argument, it is also called the \emph{chain derivative} operator. For the rest of the paper, chain differentials will always assumed to be chain derivatives and called as such.
Also, when a functional $G$ is defined as an integral with respect to a measure $\mu$ on $\mathcal{X}$ which is absolutely continuous with respect to the reference measure $\lambda$, the term $\delta G (f,\delta_x)$ will be understood as the Radon-Nikodym derivative of the measure $\mu' : B \mapsto \delta G (f,\mathds{1}_B)$ evaluated at point $x$, i.e.
\begin{equation}
\delta G (f,\delta_x) := \dfrac{\mathrm{d} \mu' }{\mathrm{d} \lambda}(x),
\end{equation}
for any appropriate function $f$ on $\mathcal{X}$ and any point $x \in \mathcal{X}$. In the context of this paper, this property holds for the \ac{pgfl} $\Gcal_\Phi$ of any point process $\Phi$ since its probability distribution $P_\Phi$ admits a density \ac{wrt} the reference measure $\lambda$. In particular,
\begin{align}
p_\Phi^{(n)}(x_1, \ldots, x_n) &= \frac{1}{n!}\delta^n\Gcal_\Phi(h;\delta_{x_1}, \ldots, \delta_{x_n})|_{h=0}, \label{eq:probaderivativedensity}
\end{align}
for any points $x_i \in \Xcal$, $1 \leq i \leq n$. This result is similar to the extraction rule (52) in \cite{Mahler2003Multitarget}, allowing for the evaluation of the multitarget density of a \ac{rfs} in the \emph{set} $\{x_1,\ldots,x_n\}$.


\section{Four relevant examples of point processes}
\label{sec:pointprocessesexamples} 
This section presents three well-established point processes in the context of multi-object estimation, namely, the \ac{iid}, Bernoulli, and Poisson point processes. It then introduces the Panjer point process and its fundamental properties.
\subsection{\acs{iid} cluster process}
An \ac{iid} cluster process with cardinality distribution $\rho$ on $\Nbb$ and spatial distribution $s$ on $\Xcal$ describes a population whose size is described by $\rho$, and whose objects' state are \ac{iid} according to $s$. Its \ac{pgfl} is given by
\begin{equation}
\Gcal_\mathrm{i.i.d.}(h) = \sum_{n\geq 0} \rho(n) \left[\int h(x)s(\d x) \right]^n .\label{eq:iid}
\end{equation}
In the construction of the \ac{cphd} filter, the predicted target process $\Phi_{k|k-1}$ is assumed \ac{iid} cluster \cite{Mahler2007PHD}.
\subsection{Bernoulli process}
A Bernoulli point process with parameter $0 \leq p \leq 1$ and spatial distribution $s$ is an \ac{iid} cluster process with spatial distribution $s$, whose size is $1$ with probability $p$ and $0$ with probability $q = (1 - p)$. Its \ac{pgfl} is given by
\begin{equation}
\label{eq:bernoulli}
\Gcal_\mathrm{Bernoulli}(h)= q + p \int h(x)s(\d x).
\end{equation}
In the context of target tracking, Bernoulli processes are commonly used to describe binary events such as the detection or survival of individual targets.
\subsection{Poisson process}
A Poisson process with parameter $\lambda$ and spatial distribution $s$ is an \ac{iid} cluster process with spatial distribution $s$, whose size is Poisson distributed with rate $\lambda$. Its \ac{pgfl} is given by
\begin{equation}
\label{eq:poisson}
\Gcal_\mathrm{Poisson}(h) =  \exp \left(\int [h(x)-1]\mu(\d x)\right),
\end{equation}
where the intensity measure $\mu$ of the process  is such that $\mu(\d x) =  \lambda s(\d x)$. Due to its simple form and its prevalence in many natural phenomena, the Poisson point process is a common and well-studied modelling choice. It can be shown that the intensity \eqref{eq:nmomentmeasure} and the variance \eqref{eq:variance} of a Poisson process are equal when evaluated in any region $B \in \Bcal(\Xcal)$, i.e., $\mu_{\Phi}(B) = \var_{\Phi}(B)$. In other words, the random variable describing the number of objects within $B$ has equal mean and variance. This property holds in particular for $B = \Xcal$. In the construction of the \ac{phd} filter, the predicted target process $\Phi_{k|k-1}$ is assumed Poisson \cite{Mahler2003Multitarget}.
\subsection{Panjer process}\label{subsubsec:panjerprocess}
A Panjer point process with parameters $\alpha$ and $\beta$ and spatial distribution $s$ is an \ac{iid} cluster process with spatial distribution $s$, whose size is Panjer distributed with parameters $\alpha$ and $\beta$ \cite{Fackler2009Panjer}, i.e., whose cardinality distribution is given by
\begin{equation}
\rho(n) = {-\alpha \choose n}\bigg(1+\frac{1}{\beta}\bigg)^{-\alpha}\bigg(\frac{-1}{\beta+1}\bigg)^n, \label{eq:panjerpmf}
\end{equation}
for any $n\in \mathbb{N}$, where either $\alpha,\beta  \in \mathbb{R}_{>0}$ or ${\alpha \in \mathbb{Z}_{<0}}$ and ${\beta \in \mathbb{R}_{<0}}$.\footnote{Note that negative, non-integer values of $\alpha$ yield complex values, and are thus discarded.} 
The particular nature of the Panjer process is determined by the values $\alpha$ and $\beta$:
\begin{itemize}
\item For finite and positive $\alpha$ and $\beta$, \eqref{eq:panjerpmf} describes a negative binomial distribution.
\item For finite and negative $\alpha$ and $\beta$ we obtain a binomial distribution.\footnote{In \cite{Fackler2009Panjer}, the binomial and negative binomial distributions are given in different forms but are equivalent to \eqref{eq:panjerpmf}.}
\item The limit case $\alpha,\beta \rightarrow \infty$ with constant ratio $\lambda=\frac{\alpha}{\beta}$ yields a Poisson process with parameter $\lambda$ \cite{klugman2012loss,Schlangen2016PHD}. 
\end{itemize}
The \ac{pgfl} of a negative binomial process is given in \cite{Daley2003Introduction}, and it can be extended to the Panjer point process as follows:
\begin{prop}
\label{prop:panjerpgfl}
The \ac{pgfl} of a Panjer process with parameters $\alpha$, $\beta$ is given by
\begin{equation}
\label{eq:panjer}
\Gcal_\mathrm{Panjer}(h) = \left( 1+\frac{1}{\beta} \int [1-h(x)]s(\d x)\right)^{-\alpha}.
\end{equation}
\end{prop}
The proof is given in the appendix. The parameters of a Panjer point process are linked to the first- and second-order moment of its cardinality distribution as follows: 
\begin{prop}
\label{prop:alphabeta}
The parameters $\alpha_{\Phi}, \beta_{\Phi}$ of a Panjer process $\Phi$ are such that
\begin{align}
\alpha_{\Phi} &= \cfrac{\mu_\Phi(\Xcal)^2}{\var_\Phi(\Xcal)-\mu_\Phi(\Xcal)},\label{eq:alpha}\\
\beta_{\Phi} &= \cfrac{\mu_\Phi(\Xcal)}{\var_\Phi(\Xcal)-\mu_\Phi(\Xcal)}.\label{eq:beta}
\end{align}
\end{prop}
The proof is given in the appendix. It can be seen from Eqs.~\eqref{eq:alpha}, \eqref{eq:beta} that binomial and negative binomial point processes have a size with \emph{larger} and \emph{smaller} variance than mean, respectively. In particular, a negative binomial point process can model a population whose size is highly uncertain, such as the clutter process in the \ac{phd} filter with negative binomial clutter \cite{Schlangen2016PHD}.

\section{The second-order \ac{phd} filter with variance in target number}
\label{sec:filter}
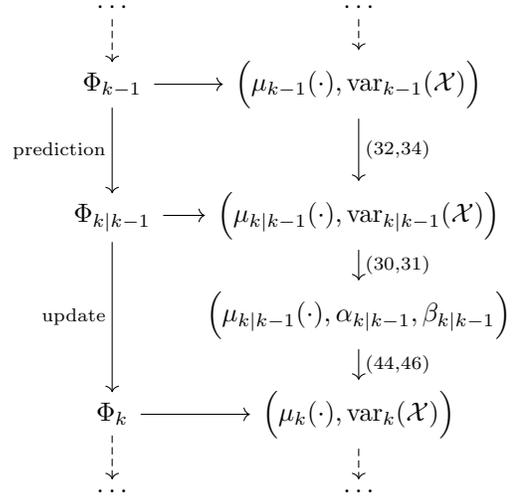
\begin{figure}
\begin{center}
\begin{tikzcd}[column sep = 1.2em, row sep = 1.2em]
 \dots \arrow[dashed]{d}   &  \dots \arrow[dashed]{d}  \\
   {\Phi}_{k-1}\arrow{dd}[swap]{\mathrm{prediction}}\arrow{r}& \Big(\mu_{k-1}(\cdot),\var_{k-1}(\Xcal)\Big)\arrow{dd}{(\ref{eq:phdpred},\ref{eq:varpred})} \\ 
   & \\
   {\Phi}_{k|k-1} \arrow{dd}[swap]{\mathrm{update}} \arrow{r}   &  \Big(\mu_{k|k-1}(\cdot), \var_{k|k-1}(\Xcal)\Big)\arrow{d}{(\ref{eq:alpha},\ref{eq:beta})}\\
   & \Big( \mu_{k|k-1}(\cdot),\alpha_{k|k-1}, \beta_{k|k-1}\Big) \arrow{d}{(\ref{eq:phdupdate},\ref{eq:varupdate})}\\
  {\Phi}_{k}\arrow[dashed]{d}\arrow{r}&\Big( \mu_{k}(\cdot), \var_{k}(\Xcal)\Big)\arrow[dashed]{d} \\
  \dots & \dots 
\end{tikzcd}
\end{center}
\caption{Data flow of the second-order \ac{phd} filter at time $k$. In addition to the intensity function $\mu$ it propagates the scalar $\var(\Xcal)$, describing the variance in the estimated number of targets in the whole state space.\label{fig:dataflow}}
\end{figure}

The intensity measure of the target process (or its density) plays an important role in the construction of multi-object filters; it is propagated by both the \ac{phd} \cite{Mahler2003Multitarget} and \ac{cphd} filters \cite{Mahler2007PHD}. The \ac{cphd} propagates also the cardinality distribution of the target process, whereas the estimated number of targets in the scene is described by the \ac{phd} filter through the mean value $\mu_{\Phi}$ only. Rather than the full cardinality  distribution, the second-order \ac{phd} filter in this section propagates the variance $\var_{\Phi}(\Xcal)$ instead. In order to do so, the Poisson or \ac{iid} cluster assumption on the predicted target process $\Phi_{k|k-1}$ is replaced by a Panjer assumption. The data flow of this filter is depicted in Fig.~\ref{fig:dataflow}.

\subsection{Time prediction step (time $k$)}
In the time prediction step, the posterior target process $\Phi_{k-1}$ is predicted to $\Phi_{k|k-1}$ based on prior knowledge on the dynamical behaviour of the targets. The assumptions of the time prediction step can be stated as follows:
\begin{assum}
\label{assum:1}
\textcolor{white}{.}
\begin{enumerate}[(a)]
\item The targets evolve independently from each other;
\item A target with state $x \in \Xcal$ at time $k-1$ survived to the current time $k$ with probability $p_{\s,k}(x)$; if it did so, its state evolved according to a Markov transition kernel $t_{k|k-1}(\cdot|x)$;
\item New targets entered the scene between time $k-1$ and $k$, independently of the existing targets and described by a newborn point process $\Phi_{\b, k}$ with \ac{pgfl} $\Gcal_{\b, k}$.
\end{enumerate}
\end{assum}
\begin{assum}
\label{assum:1b}
\textcolor{white}{.}
\begin{enumerate}[(a)]
\item The probability of survival is uniform over the state space, i.e., $p_{\s, k}(x) := p_{\s, k}$ for any $x \in \Xcal$.
\end{enumerate}
\end{assum}
Note that Assumptions \ref{assum:1} are those of the original \ac{phd} filter; in particular, the second-order \ac{phd} filter does not require a specific form for the posterior process $\Phi_{k-1}$ or the birth process $\Phi_{\b, k}$. 

\begin{thm}[Intensity prediction \cite{Mahler2003Multitarget}]
\label{thm:phdpred}
Under Assumptions \ref{assum:1}, the intensity measure $\mu_{k|k-1}$ of the predicted target process $\Phi_{k|k-1}$ is given by 
\begin{equation}
\label{eq:phdpred}
\mu_{k|k-1}(B) = \mu_{\b, k}(B) + \mu_{\s,k}(B),
\end{equation}
in any $B \in \Bcal(\Xcal)$, where $\mu_{\s,k}$ is the intensity measure of the process describing the surviving targets
\begin{equation}
\label{eq:phdpredsurv}
 \mu_{\mathrm{s},k}(B) = \int p_{\s, k}(x)t_{k|k-1}(B|x) \mu_{k-1}(\d x),
\end{equation}
and $\mu_{\b,k}$ is the intensity measure of the newborn process $\Phi_{\b, k}$.
\end{thm}

\begin{thm}[Variance prediction]
\label{thm:varpred}
Under Assumptions \ref{assum:1}, the variance $\var_{k|k-1}$ of the predicted target process $\Phi_{k|k-1}$ is given by
\begin{equation}
 \label{eq:varpred}
\var_{k|k-1}(B) = \var_{\b, k}(B) + \var_{\s, k}(B),
\end{equation}
in any $B \in \Bcal(\Xcal)$, where $\var_{\s, k}$ is the variance of the process describing the surviving targets
\begin{align}
\label{eq:var_s}
  &\var_{\s, k}(B) = \mu_{\mathrm{s},k}(B) \Big[1 - \mu_{\mathrm{s},k}(B)\Big]\nonumber
  \\
  &+\int p_{\s, k}(x)p_{\s, k}(x')t_{k|k-1}(B|x)t_{k|k-1}(B|x') \nu_{k-1}^{(2)}(\d(x,x')),
\end{align}
and $\var_{\b, k}$ is the variance of the newborn process $\Phi_{\b, k}$. 
\end{thm}
The proof is given in the appendix. Note that the propagation of the regional variance \eqref{eq:varpred} -- i.e., the variance $\var_{k|k-1}(B)$ in any $B \in \Bcal(\Xcal)$ -- requires the posterior second-order factorial moment $\nu_{k-1}^{(2)}$, which is not available from the posterior information $\mu_{k-1}(\cdot), \var_{k-1}(\Xcal)$ (see data flow in Fig.~\ref{fig:dataflow}). However,  considering the additional Assumption~\ref{assum:1b}, the variance of the predicted target process $\Phi_{k|k-1}$ evaluated in the whole state space becomes
\begin{cor}[Variance prediction, uniform $p_{\s,k}$]
\label{cor:varpred}
Under Assumptions \ref{assum:1} and  \ref{assum:1b}, the variance $\var_{k|k-1}$ of the predicted target process $\Phi_{k|k-1}$ evaluated in the whole state space $\Xcal$ is given by
\begin{equation}
 \label{eq:varpredassumption}
\var_{k|k-1}(\Xcal) = \var_{\b, k}(\Xcal) + \var_{\s, k}(\Xcal),
\end{equation}
where $\var_{\s, k}$ is the variance of the process describing the surviving targets
\begin{equation}
\label{eq:var_s_assumption}
 \var_{\s, k}(\Xcal) = p_{\s,k}^2\var_{k-1}(\Xcal) + p_{\s,k}[1- p_{\s, k}]\mu_{k-1}(\Xcal),
\end{equation}
and $\var_{\b, k}$ is the variance of the newborn process $\Phi_{\b, k}$. 
\end{cor}
The proof is given in the appendix. The results in Thm.~\ref{thm:phdpred} and Cor.~\ref{cor:varpred} produce the predicted quantities $\mu_{k|k-1}$, $\var_{k|k-1}(\Xcal)$ from their posterior values $\mu_{k-1}$, $\var_{k-1}(\Xcal)$.

\subsection{Data update step (time $k$)} \label{subsec:panjerupdate}
In the data update step, the predicted process $\Phi_{k|k-1}$ is updated to $\Phi_k$ given the current measurement set $Z_k$, collected from the sensor. The date update step relies on the following assumptions:
\begin{assum}
\label{assum:2}
\textcolor{white}{.}
\begin{enumerate}[(a)]
\item The predicted target process $\Phi_{k|k-1}$ is Panjer, with parameters $\alpha_{k|k-1}$, $\beta_{k|k-1}$ and spatial distribution $s_{k|k-1}$.
\item The measurements originating from target detections are generated independently from each other.
\item A target with state $x \in \Xcal$ is detected with probability $p_{\d,k}(x)$; if so, it produces a measurement whose state is distributed according to a likelihood $l_k(\cdot|x)$.
\item The clutter process, describing the false alarms produced by the sensor, is Panjer with parameters $\alpha_{\c, k}$, $\beta_{\c, k}$ and spatial distribution $\s_{\c,k}$.
\end{enumerate}
\end{assum}
Before stating the data update equations for the second-order \ac{phd} filter, recall the \emph{Pochhammer symbol} or \emph{rising factorial} $(\zeta)_n$ for any $\zeta\in \mathbb{R}$ and $n\in\mathbb{N}$:
\begin{equation}
\label{eq:pochhammer}
(\zeta)_n := \zeta(\zeta+1)\cdots (\zeta+n-1),\quad(\zeta)_0 :=1.
\end{equation}
Following the notations used in \cite{Delande2014Regional} and introduced in \cite{Vo2007Analytic}, we define the corrective terms
\begin{equation}
\begin{split}
&\ell_{u}(z) := \cfrac{\Upsilon_{u}(Z_k \backslash \{z\})}{\Upsilon_{0}(Z_k)},\quad \ell_u(\phi) := \cfrac{\Upsilon_{u}(Z_k)}{\Upsilon_{0}(Z_k)},
\end{split}
\label{eq:corrective1}
\end{equation}
for any $u \in \Nbb$ and any $z \in Z_k$, where
\begin{equation}
\label{eq:ypsilon}
\begin{split}
\Upsilon_{u}(Z) := \sum_{j=0}^{|Z|} \frac{(\alpha_{k|k-1})_{j+u}}{(\beta_{k|k-1})^{j+u}} \frac{(\alpha_{\c, k})_{|Z|-j}}{(\beta_{\c, k} + 1)^{|Z|-j}}F_\d^{-j-u}e_{j}(Z),
\end{split}
\end{equation}
for any $Z \subseteq Z_k$, where $F_\d$ is the scalar given by
\begin{equation}
F_\d := \int \left[1 + \frac{p_{\d, k}({x})}{\beta_{k|k-1}}\right]\mu_{k|k-1}(\d{x}),
\end{equation}
and $e_j$ is the $j$-th elementary symmetric function
\begin{equation}
e_j(Z) := \sum_{\substack{{Z'}\subseteq Z\\|{Z'}|=j}}\prod_{z\in {Z'}}\frac{\mu_k^z(\Xcal)}{s_{\c, k}(z)},
\label{eq:esf}
\end{equation}
with 
\begin{equation}
\label{eq:mu_z}
\mu_{k}^z(B) =  \int_B p_{\d, k}(x) l_k(z|x) \mu_{k|k-1}(\d x),
\end{equation}
for any $B \in \Bcal(\Xcal)$.\footnote{In these definitions, the time subscripts on the $\ell_u$, $\Upsilon_u$, $F_{\d}$, and $e_j$ terms are omitted for the sake of simplicity.}
\begin{thm}[Intensity update]
\label{thm:phdupdate} Under Assumptions~\ref{assum:2}, the intensity measure $\mu_k$ of the updated target process $\Phi_k$ is given by
\begin{equation}
\label{eq:phdupdate}
\begin{split}
\mu_k(B) = \mu_{k}^\phi(B)\ell_1(\phi) + \sum_{z\in Z_k} \frac{\mu_{k}^z(B)}{s_{\c, k}(z)} \ell_{1}(z),
\end{split}
\end{equation}
in any $B \in \Bcal(\Xcal)$, where the missed detection term $\mu_{k}^\phi$ is given by
\begin{equation}
\label{eq:mu_phi}
\mu_{k}^\phi(B)= \int_B (1 - p_{\d, k}(x)) \mu_{k|k-1}(\d x).
\end{equation} 
\end{thm}
\begin{thm}[Variance update]
\label{thm:varupdate}
Under Assumptions~\ref{assum:2}, the variance $\var_k$ of the updated target process $\Phi_k$ is given by
\begin{equation}
\label{eq:varupdate}
\begin{split}
&\var_k(B)=\mu_k(B)
+\mu_k^\phi(B)^2\left[\ell_2(\phi)-\ell_1(\phi)^2\right] \\
&+2\mu_k^\phi(B)\sum_{z\in Z_k}~\frac{\mu_k^z(B)}{s_{\c,k}(z)}\left[\ell_{2}(z)-\ell_1(\phi)\ell_{1}(z)\right]\\
&+\sum_{z,z'\in Z_k}~\frac{\mu_k^z(B)}{s_{\c,k}(z)}\frac{\mu_k^{z'}(B)}{s_{\c,k}(z')}\left[\ell_{2}^{\neq}(z,z')-\ell_{1}(z)\ell_{1}(z')\right],
\end{split}
\end{equation}
in any $B \in \Bcal(\Xcal)$, with
\begin{equation}
\ell_{2}^{\neq}(z,z') := \left\{ \begin{array}{ll} 
 \cfrac{\Upsilon_{2}(Z_k \backslash \{z,z'\})}{\Upsilon_{0}(Z_k)}, & z\neq z', \\
 0, & \mathrm{otherwise}.
\end{array}\right.
\label{eq:corrective2}
\end{equation}
\end{thm}
The proofs of Thms~\ref{thm:phdupdate} and \ref{thm:varupdate} are given in the appendix. Together with Eqs~\eqref{eq:alpha}, \eqref{eq:beta}, the results in Thms~\ref{thm:phdupdate}, \ref{thm:varupdate} produce the updated quantities $\mu_k$, $\var_k(\Xcal)$ from their predicted values $\mu_{k|k-1}$, $\var_{k|k-1}(\Xcal)$.

As mentioned earlier in Sec.~\ref{subsubsec:panjerprocess}, a Panjer distribution converges to a Poisson distribution for suitable parameters $\alpha$, $\beta$. An interesting consequence for the intensity update of the second-order \ac{phd} filter proposed in Eq.~\eqref{eq:phdupdate} is that

\begin{cor}[Intensity update: limit cases]
\label{cor:limit}

If, in addition to Assumptions~\ref{assum:2}, the predicted point process $\Phi_{k|k-1}$ is assumed Poisson, i.e., $\alpha_{k|k-1}, \beta_{k|k-1} \rightarrow \infty$ with constant ratio $\lambda_{k|k-1} = \frac{\alpha_{k|k-1}}{\beta_{k|k-1}}$, then the intensity update \eqref{eq:phdupdate} converges to the intensity update of the \ac{phd} filter with Panjer clutter given in \cite{Schlangen2016PHD}.

Furthermore, if the clutter process is assumed Poisson as well, i.e., $\alpha_{\c, k}, \beta_{\c, k} \rightarrow \infty$ with constant ratio $\lambda_{\c, k} = \frac{\alpha_{\c, k}}{\beta_{\c, k}}$, then the intensity update given in \cite{Schlangen2016PHD} converges to the intensity update of the original \ac{phd} filter \cite{Mahler2003Multitarget}.
\end{cor}

With Cor.~\ref{cor:limit}, the second-order \ac{phd} filter presented in this paper can be seen as a generalisation of the original \ac{phd} filter\footnote{Under the proviso that the additional assumption \ref{assum:1b} is met, i.e., the probability of survival $p_{\s, k}$ is uniform over the state space.}. Note that the expression of the intensity \eqref{eq:phdupdate} and update \eqref{eq:varupdate} of the updated target process are remarkably similar to their counterpart in the \ac{cphd} filter, and only differ on the expressions of the corrective terms $\ell_u$ \cite{Delande2014Regional}. Both filters involve the computation of elementary symmetric functions $e_{j}(Z)$ on subsets $Z$ of the measurement set $Z_k$. Each function has a computational cost of $\mathcal{O}(|Z|\log^2|Z|)$ \cite{Vo2007Analytic}; the \ac{cphd} requires the computation for sets of the form $Z_k$, and $Z_k\setminus\{z\}$, for a total cost of $\mathcal{O}(|Z_k|^2\log^2|Z_k|)$, while the proposed solution requires the computation for sets of the form $Z_k$, $Z_k\setminus\{z\}$, $Z_k\setminus\{z, z'\}$, for a total cost of $\mathcal{O}(|Z_k|^3\log^2|Z_k|)$. However, while the \ac{cphd} filter requires the computation of the $\Upsilon^u(n)$ terms \cite[Eq.~(14)]{Vo2007Analytic} for each possible target number $n$ (to a maximum number of targets $N_{\max}$, set as a parameter), the proposed filter requires the computation of the $\Upsilon_u$ terms \eqref{eq:ypsilon} \emph{only once}. The complexity of the proposed filter is thus significantly lower than for the \ac{cphd} filter, as it will be illustrated in the simulation results in Sec.~\ref{sec:experiments}, and the difference in complexity increases with the value of the \ac{cphd} parameter~$N_{\max}$.

\section{Regional correlations for \ac{phd} filters}
\label{sec:correlation}
In order to assess the mutual influence of the estimated number of targets in two regions $B,B'\in \Bcal(\Xcal)$, we compute in this section the statistical correlation \eqref{eq:correlation} of the updated target process $\Phi_k$ for the \ac{phd}, second-order \ac{phd} and \ac{cphd} filters.  

\begin{prop}[Covariance of the \ac{phd} filters] \textcolor{white}{.}\\
\label{prop:covariance}
\vspace{2pt}Let ${B,B'\in \Bcal(\Xcal)}$ be two arbitrary regions in the state space. %
\\(a) \ac{phd} filter: \\Let $\lambda_{\c,k}$ be the Poisson clutter rate at time $k$. The covariance of the updated target process $\Phi_k$ in $B$, $B'$ is 
      \begin{equation}
	\begin{split}
	  &\cov_k(B \times B') = \mu_k(B \cap B') - \!\sum_{z \in Z_k} \frac{\mu^{z}_{k}(B)\mu^{z}_{k}(B')}{\big[\mu^z_{k}(\Xcal) + \lambda_{\c, k}s_{\c, k}(z)\big]^2}. \label{eq:phd_cov}
	\end{split}      
      \end{equation}
(b) Second-order \ac{phd} filter:\\ The covariance of the updated target process $\Phi_k$ in $B$, $B'$ is 
      \begin{equation}
	\begin{split}
	  &\cov_k(B \times B') 
	  \\
	  &= \mu_k(B \cap B') + \mu^{\phi}_{k}(B)\mu^{\phi}_{k}(B')[\ell_2(\phi) - \ell_1(\phi)^2] 
	  \\
	  &+ \sum_{z \in Z_k}\left[\mu^{\phi}_{k}(B) \frac{\mu^{z}_{k}(B')}{s_{\c, k}(z)} + \mu^{\phi}_{k}(B')\frac{\mu^{z}_{k}(B)}{s_{\c, k}(z)}\right]\left[\ell_2(z) - \ell_1(z)\ell_1(\phi)\right] 
	  \\
	  &+ \sum_{z, z' \in Z_k}\left[\frac{\mu^{z}_{k}(B)}{s_{\c, k}(z)}\frac{\mu^{z'}_{k}(B')}{s_{\c, k}(z')}\right]\left[\ell^{\neq}_2(z, z') - \ell_1(z)\ell_1(z')\right]. \label{eq:panjer_cov}
	\end{split}      
      \end{equation}
(c) \ac{cphd} filter:\\  The covariance of the updated target process $\Phi_k$ in $B$, $B'$ is given by \eqref{eq:panjer_cov}, where the corrective terms $\ell_1$, $\ell_2$ and $\ell_2^{\neq}$ are replaced by the values in Eqns (20), (30) of \cite{Delande2014Regional}.
\end{prop}
The proof is given in the appendix. The correlations $\corr_\Phi(B,B')$ are a direct consequence of Eq.~\eqref{eq:correlation}, using the regional variance stated in Eqns (35), (33) \cite{Delande2014Regional} for the \ac{phd} and \ac{cphd} filters and the regional variance \eqref{eq:varupdate} for the second-order \ac{phd} filter.

\section{Experiments}
\label{sec:experiments}
A \ac{gm} implementation \cite{Vo2006Gaussian,Vo2007Analytic} was used for all algorithms to make them comparable. For the \ac{cphd} filter, the maximum number of targets $N_{\max}$ is set to $150$ for all experiments. The \ac{ospa} metric per time step \cite{Schuhmacher2008Consistent} is used with the Euclidean distance (i.e.~$p=2$) and the cutoff $c=100$.

\subsection{Scenario 1}
\label{subsec:birthdeathex}
This scenario examines the robustness of the \ac{phd}, \ac{cphd}, and Panjer filters to large variations in the number of targets and focuses on a single time step when the change in target number occurs.

The size of the surveillance scene is $\SI{50}{\metre} \times \SI{50}{\metre}$. The generation of new objects is restricted to the centre of the image to prevent the objects from leaving the scene before the last time step. Their movement is generated using a nearly constant velocity model where the standard deviation of the acceleration noise is $\SI{0.3}{\metre\per\square\second}$ and the initial velocity is Gaussian normal distributed with mean 0 and standard deviation $\SI{0.5}{\metre\per\second}$ along each dimension of the state space. False alarms are generated according to a Poisson point process with uniform spatial distribution and clutter rate $\mu_\c=5$ for experiments 1.1, 1.2 and $\mu_\c=20$ for experiment 1.3. The probabilities of detection and survival are constant and set to $0.9$ and $0.99$, respectively.

\begin{enumerate}
\item[1.1] 50 targets are created in the first time step and propagated until time step 15 to give the algorithms time to settle. At time 15, the number of targets suddenly changes, either by removing some or all of the current targets without creating new objects or by creating up to 50 births while maintaining the old targets. The birth model  is Poisson with uniform spatial distribution and birth rate $\mu_\b=25$, for the three filters.
\item[1.2] The parameters are identical to experiment 1.1, except that the birth model is  negative binomial with $\mu_\b = 25$ and $\var_\b = 100$ for the Panjer and CPHD filter.
\item[1.3]
Here, only one target is created in the beginning and maintained up to time 15. At this time, from 0 to 100 targets are spontaneously created in the scene. The birth model is a negative binomial point process with uniform spatial distribution, mean $\mu_\b = 1$ and $\var_\b = 100$ for the three filters, though the \ac{phd} filter cannot exploit the information on the variance. 
\end{enumerate}
Fig.~\ref{fig:birth_death_scenario} depicts the results of this scenario. In experiment 1.1 and 1.2, the three filters estimate target birth more accurately than target death since the high survival probability,  together with a high birth rate, does not account for severe drops in the number of targets. In particular, the \ac{cphd} filter lacks flexibility and fails at recognising unexpected drops in the number of targets. Choosing negative binomial birth model allows for larger uncertainty in the number of targets and improves the quality of the estimate for the \ac{cphd} and Panjer filters. Furthermore, the variance of the Panjer filter is lower than that of the \ac{phd} filter.

Experiment 1.3 highlights a limitation of the \ac{phd} filter, which reduces the prior information on the number of newborn targets to its mean value. The \ac{cphd} and Panjer filters, on the other hand, can exploit a birth process with high variability in target number -- i.e., through a negative binomial process with large variance in target number --  in order to cope with a burst of target births. Fig.~\ref{fig:bigbirth_mean} suggests that the birth and false alarm processes are competing in the \ac{cphd} and Panjer filters when there is a significant influx in the number of newborn targets,  resulting in an offset linked to the mean number of false alarms (recall that $\mu_\c = 20$ in this case). The \ac{phd} filter, on the other hand, is unable to cope with a influx that is well beyond the Poisson model.

Average run times are omitted for this scenario as they change greatly with the different changes in target number and are therefore not very meaningful. The following scenarios will provide a more valuable insight in the computational performance.

\subsection{Scenario 2}
\label{subsec:stepex}
This scenario examines the behaviours of the \ac{phd}, \ac{cphd} and Panjer filters under the influence of increasing amounts of target birth and death.

The size of the surveillance scene is $\SI{50}{\metre} \times \SI{50}{\metre}$. The number of targets is designed to follow a stair pattern starting with $5$ initial targets, and increasing the cardinality by $10$, $15$, $20$ and $25$ targets every ten time steps until time $40$. From time $50$ onwards up to time $90$, the number of targets is decreased in reverse order, i.e.~every ten time steps, the target population is reduced by $25$, $20$, $15$, and $10$ targets. The generation of new objects is restricted to the centre of the image to prevent the objects from leaving the scene before the last time step. Their movement is generated using a nearly constant velocity model where the standard deviation of the acceleration noise is $\SI{0.1}{\metre\per\square\second}$ and the initial velocity is Gaussian normal distributed with mean 0 and standard deviation $\SI{0.3}{\metre\per\second}$ along each dimension of the state space.

From the ground truth obtained as above, measurements are created with a constant probability of detection. For comparison, two different values are chosen, i.e.~$p_\d = 0.95$ in the first experiment and $p_\d=0.6$ in the second. Each detection is corrupted with white noise with standard deviation \SI{0.2}{\metre} in each dimension. Additionally, false alarms are generated according to a Poisson point process with uniform spatial distribution and clutter rate $\mu_\c=15$. 

The three filters are parametrised with the simulation parameters above. In addition, the probability of survival is set to $p_\s = 0.98$, and target birth is modelled using a negative binomial process with uniform spatial distribution, mean $\mu_\b(\Xcal) = 1$ and variance $\var_b(\Xcal) = 100$ to account for the big changes in the number of objects. 
Each experiment on $100$ \ac{mc} runs. 

In Fig.~\ref{fig:step_gt}, an example run of  the first experiment is depicted. Fig.~\ref{fig:step} shows the estimated means and variances for all filters and all experiments over time (left column), along with the mean and standard deviation of the respective \ac{ospa} distances over time (right column).

The first experiment (Fig.~\ref{fig:NB_card}-\ref{fig:NB_ospa}) demonstrates that the three filters show a delay in the adjustment of the cardinality estimate when the population is growing, resulting in spikes of \ac{ospa} error. In general, the \ac{cphd} filter is closest to the true target number, however in case of target death, the \ac{phd} and Panjer filters prove to be more reactive despite setting the survival rate to $98\%$.

In the second experiment (cf.~Fig.~\ref{fig:pd_card}-\ref{fig:pd_ospa}), all three filters show a significant increase in the estimated variance in cardinality since target death and missed detections are hard to distinguish and therefore more missed detections lead to increased uncertainty in the number of targets. In terms of the estimated mean, on the other hand, the proposed method shows the highest reactivity to target birth and especially to target death, estimated poorly with the \ac{cphd} filter.

Table \ref{tab:runtimes} shows the averaged run time for both cases of this scenario. The prediction runs approximately $100$ times slower for the \ac{cphd} than for the first- and second-order \ac{phd} filters; this is to be expected since the complexity of the former grows proportional to the range of cardinalities for which the cardinality distribution is estimated. The update performance, on the other hand, varies greatly for different probabilities of detection: if $p_{\d}$ is low, the weight for miss-detected objects does not plummet directly and therefore the information about dead tracks is kept and propagated for longer.

\subsection{Scenario 3}
\label{subsec:spookyex}
This scenario assesses the spooky effect of the \ac{phd}, \ac{cphd}, and Panjer filters through the regional covariance introduced in this paper.

Two completely separate regions of interest, henceforth called $A$ and $B$, are depicted in Fig.~\ref{fig:scenario}. Both regions are of size $\SI{50}{\metre} \times \SI{50}{\metre}$, and they are $\SI{100}{\metre}$ apart horizontally. In each region, $10$ targets are initialised in the first time step and they survive throughout $100$ time steps. Again, the generation of new objects is restricted to the centre of each region to prevent the objects from leaving the scene before the last time step. Their movement is generated using a nearly constant velocity model where the standard deviation of the acceleration noise is $\SI{0.1}{\metre\per\square\second}$ and the initial velocity is Gaussian normal distributed with mean 0 and standard deviation $\SI{0.3}{\metre\per\second}$ along each dimension of the state space.

Measurements are created with the (constant) probability of detection $p_\d = 0.9$. Each detection is corrupted with white noise with standard deviation \SI{0.2}{\metre} in each dimension. Additionally, false alarms are generated in each region according to a Poisson point process with uniform spatial distribution (in the region) and clutter rate $\mu_\c(A) = \mu_\c(B) =20$. 

The three filters are parametrised with the simulation parameters above. In addition, the probability of survival is set to $p_\s = 0.98$, and target birth is modelled using a negative binomial point process with uniform spatial distribution (in the region) with mean $\mu_\b(\Xcal) = 1$ and variance $\var_b(\Xcal) = 100$ to account for sudden changes in the number of objects. 

In order to analyse the spooky effect on this scenario, \emph{all} objects in region $B$ are forced to be miss-detected every $10$ time steps, additionally to the modelled natural missed detections in the scene. Fig.~\ref{fig:card_phd}-\ref{fig:card_cphd} show the estimated regional means and regional variances for the three filters in both regions. In case of the \ac{phd} filter (cf.~Fig.~\ref{fig:card_phd}), the intensity in region $A$ is unaffected by the sudden drop in the intensity in region $B$. The proposed filter, in contrast, reacts with a slight drop in the intensity of region $A$ when the targets in $B$ are missed, and it compensates sightly in each subsequent time step (Fig.~\ref{fig:card_panjer}). The biggest effect by far is noticed with the \ac{cphd} filter, as seen in Fig.~\ref{fig:card_cphd}. Every time the objects in $B$ stay undetected, the intensity in that region does not drop as low as for the other two filters, but the intensity in region $A$ increases notably to approximately $12$ targets.

The observed behaviour can be further illustrated by looking at the correlation of $A$ and $B$ under the \ac{phd}, Panjer and \ac{cphd} filters, exploiting the covariance of the three filters given in Sec.~\ref{sec:correlation}. Eq.~\eqref{eq:phd_cov} shows that the covariance of the \ac{phd} filter is 0 if the two regions are disjoint and the region of origin of each measurement is unambiguous; this is clearly seen in the correlation depicted in Fig.~\ref{fig:correlation}. The same figure shows a strongly negative correlation in the case of the \ac{cphd} filter, which highlights the spooky effect: the filter compensates for the lost intensity mass in region $B$ by introducing it in region $A$. The Panjer filter shows a milder but positive correlation, as the sudden drop/increase in intensity mass in region $B$ goes along with a smaller drop/increase in region $A$. These results suggest that, on these experiments, the Panjer filter exhibits a milder spooky effect than the \ac{cphd} filter.

Table \ref{tab:runtimes} shows the averaged run time for this scenario, showing a coherent image with the findings above.

\begin{center}

\begin{tabular}{| c | c | r r r | }

\hline

& scenario & \ac{phd}\hspace{1.2mm} & Panjer\hspace{0.5mm} & \ac{cphd} \\ \hline

\parbox[t]{2mm}{\multirow{3}{*}{\rotatebox[origin=c]{90}{Pred.}}}& 2.1 &0.0143  &  0.0150  &  0.9761 \\

& 2.2 & 0.0266  &  0.0285  &  1.0901 \\

& 3 & 0.0121  &  0.0143  &  0.6734  \\

\hline

\parbox[t]{2mm}{\multirow{3}{*}{\rotatebox[origin=c]{90}{Update}}}& 2.1 &     3.9233  &  6.2693  & 23.0930 \\

& 2.2 & 36.6506 &  40.9254 & 46.9830 \\

& 3  &  2.1956  &  2.3640  & 10.3355 \\

\hline
\end{tabular}

\captionof{table}{Runtimes for experiments 2 and 3, averaged over all time steps and Monte Carlo runs. The times are given in seconds.\label{tab:runtimes}}

\end{center}

\section{Conclusion}
\label{sec:conclusion}
A new second-order \ac{phd} filter has been introduced, propagating the variance in the estimated number of targets alongside the first-order moment of the target process. The Panjer point process is introduced in order to approximate the multi-target predicted process and to model the false alarm process. Described with two parameters, a Panjer distribution encompasses the binomial, Poisson, and negative binomial distribution; the resulting second-order \ac{phd} filter provides more flexibility in the modelling phase than the \ac{phd} filter. The proposed filter is implemented with a Gaussian mixture algorithm, and compared to the \ac{phd} and \ac{cphd} filters on simulated data where it proved to be more robust to changes in the number of targets of unusually large extent. In a more usual scenario, the three filters showed similar performance; the proposed filter proved more reactive to the disappearance of targets than the \ac{cphd} filter, while having a significantly lower computational complexity.

The regional covariance of a point process is introduced in order to analyse the correlation between the estimated number of targets in disjoint regions of the state space, and to assess quantitatively the well-known spooky effect of the three filters on a simulated scenario. The results showed that the estimated targets in the two regions were uncorrelated with the \ac{phd} filter, strongly negatively correlated with the \ac{cphd} filter, and midlly positively correlated with the proposed second-order \ac{phd} filter.


\section*{Appendix A: Proofs}
The appendix provides the proofs for the results in Sec.~\ref{sec:pointprocessesexamples} and \ref{sec:filter}. 

\subsection{Differentiation rules}
We first introduce the following differentiation rules, whose proofs are given in \cite{Schlangen2016PHD}.
\begin{lem}
\label{lem:derivatives}
Let $G$ be a linear functional.
\begin{enumerate}[(a)]
\item The $n$th order derivative of the composition $\exp(G(h))$ can be written as
\begin{equation}
\label{eq:exp_derivativen}
\delta^n (\exp \circ G)(h;\eta_1,\dots,\eta_n) = \exp(G(h)) \prod_{i=1}^n \delta G(h;\eta_i) .
\end{equation}
\item The $n$th order derivative of the composition $(G(h))^{-\alpha}$ is derived to be
\begin{equation}
\begin{split}
\label{eq:alpha_derivativen}
\delta^n& (G^{-\alpha})(h;\eta_1,\dots,\eta_n) \\
&= (-1)^n(\alpha)_n~G(h)^{-\alpha-n} \prod_{i=1}^n \delta G(h;\eta_i)  
\end{split}
\end{equation}
with $(\cdot)_n$ being the Pochhammer symbol \eqref{eq:pochhammer}.
\end{enumerate}
\end{lem}

\subsection{Proof of Prop.~\ref{prop:panjerpgfl}}
\begin{proof}
Since a Panjer point process is an \ac{iid} point process, let us start with equation \eqref{eq:iid}, inserting \eqref{eq:panjerpmf} for $\rho$:
\begin{subequations}
\begin{align}
&\Gcal_\mathrm{Panjer}(h) \nonumber \\ 
&\stackrel{\eqref{eq:iid}}{=}\sum_{n\geq 0} {-\alpha \choose n}\bigg(1+\frac{1}{\beta}\bigg)^{-\alpha}\bigg(\frac{-1}{\beta+1}\bigg)^n\left[ \int h(x)s(\d x) \right]^n\\
&=\bigg(1+\frac{1}{\beta}\bigg)^{-\alpha}\hspace{-0.1cm}\sum_{n\geq 0}{-\alpha \choose n}\left[\frac{-1}{\beta+1} \int  h(x)s(\d x) \right]^n\\
&=\bigg(1+\frac{1}{\beta} \bigg)^{-\alpha}\left[1- \frac{1}{\beta+1} \int  h(x)s(\d x) \right]^{-\alpha}\label{subeq:binomialseries}\\
&=\left[1+\frac{1}{\beta}- \frac{1}{\beta} \int  h(x)s(\d x) \right]^{-\alpha}\\
&=\left[1+\frac{1}{\beta} \int [1- h(x)]s(\d x) \right]^{-\alpha}
\end{align}
\end{subequations}
Equality \eqref{subeq:binomialseries} follows from the binomial series.
\end{proof}

\subsection{Proof of Prop.~\ref{prop:alphabeta}}
\begin{proof}
Let us derive the mean and variance of a Panjer process with parameters $\alpha, \beta$ and spatial distribution $s$, for arbitrary regions $B, B \in \Bcal(\Xcal)$:
\begin{subequations}
\label{eq:mean_der}
\begin{align}
&\mu(B) \stackrel{\eqref{eq:muderivative}}{=} \delta \Gcal_\mathrm{Panjer}(h;\mathds{1}_B)\bigg|_{h=1}\\ 
&\stackrel{\eqref{eq:panjer}}{=}\delta\left(\left[ 1+\frac{1}{\beta} \int [1-h(x)]s(\d x)\right]^{-\alpha};\mathds{1}_B\right)\bigg|_{h=1}\\
&\stackrel{\eqref{eq:alpha_derivativen}}{=}-\alpha\hspace{-0.1cm}\left[ 1+\frac{1}{\beta} \int [1-1]s(\d x)\right]^{-\alpha-1}\hspace{-0.1cm}\left[-\frac{1}{\beta}\int_B s(\d x) \right]\\
&=\frac{\alpha}{\beta} \int_B s(\d x).
\end{align}
\end{subequations}
\begin{subequations}
\begin{align}
 &\mu^{(2)}(B\times B') = \delta^2 \Gcal_\mathrm{Panjer}(e^{-f};\mathds{1}_B,\mathds{1}_{B'})\bigg|_{f=0} \\
 &= \frac{(\alpha)_2}{\beta^2}\left[ 1+\frac{1}{\beta} \int [1-e^0]s(\d x)\right]^{-\alpha-2}\int_Be^0 s(\d x) \int_{B'}e^0 s(\d x)\nonumber\\
 &+ \frac{\alpha}{\beta}\left[ 1+\frac{1}{\beta} \int [1-e^0]s(\d x)\right]^{-\alpha-1} \int_{B\cap B'} s(\d x)\\
  &= \frac{(\alpha)_2}{\beta^2}\int_B s(\d x) \int_{B'} s(\d x)+ \frac{\alpha}{\beta} \int_{B\cap B'} s(\d x).
\end{align}
\end{subequations}
Therefore,
\begin{subequations}
\label{eq:var_der}
\begin{align}
\var(B) &\stackrel{\eqref{eq:variance}}{=} \mu^{(2)}(B\times B) - \left[ \mu(B) \right]^2 \\
&=\mu(B) \left( 1+\frac{1}{\beta} \int_B s(\d x)\right).
\end{align}
\end{subequations}
From \eqref{eq:mean_der} and \eqref{eq:var_der} we get
\begin{equation}
  \left\{
    \begin{aligned}
      \mu(\Xcal) &= \frac{\alpha}{\beta}
      \\
      \var(\Xcal) &= \mu(\Xcal)\left(1 + \frac{1}{\beta}\right),
    \end{aligned}
  \right.
\end{equation}
which yields the desired result when solved for $\alpha$ and $\beta$.
\end{proof}

\subsection{Proof of Thm.~\ref{thm:varpred}}
\begin{proof}
In the following, we denote by $\Gcal_{\s, k}$ the \ac{pgfl} of the point process describing the evolution of a target from the previous time step, which might have survived (or not) to the present time step. For the sake of simplicity, we shall omit the time subscripts on the quantities related to the survival and birth process.
 
The first step of the proof is to formulate the \ac{pgfl} of the prediction process. In order to determine the variance as formulated in Eq.~\eqref{eq:variance}, the second-order moment of the \ac{pgfl} has to be computed and the square of the predicted intensity \eqref{eq:phdpred} be subtracted from the result. The second-order moment will lead to four terms that are computed separately. The \ac{pgfl} $\Gcal_{k|k-1}$ of the predicted target process takes the form 
\begin{equation}
\label{eq:predpgfl}
\Gcal_{k|k-1}(h) = \Gcal_\b(h)\Gcal_{k-1}(\Gcal_\s(h|\cdot)).
\end{equation}
Here, the multiplicative structure stems from the independence between the newborn targets and those surviving from the previous time step; the composition appears because the survival process applies to each preexisting target described by the updated target process $\Phi_{k-1}$ from the previous time step \cite[Eq.~5.5.18]{Daley2003Introduction}.

In order to produce the variance $\var_{k|k-1}$ of the predicted process via \eqref{eq:variance} we first build the second-order moment $\mu^{(2)}_{k|k-1}(B\times B')$ in arbitrary regions $B, B' \in \Bcal(\Xcal)$. From \eqref{eq:muderivative} we have
\begin{subequations}
\label{eq:secondordertotal}
\begin{align}
&\mu^{(2)}_{k|k-1}(B\times B') = \delta^2 \Lcal_{k|k-1}(f;\mathds{1}_B,\mathds{1}_{B'})\big|_{f=0}
\\
&= \delta^2 \Gcal_{k|k-1}(e^{-f};\mathds{1}_B,\mathds{1}_{B'})\big|_{f=0}.
\intertext{The product rule \eqref{eq:productrule} gives}
&\mu^{(2)}_{k|k-1}(B\times B') = \delta^2 \Gcal_\b(e^{-f};\mathds{1}_B;\mathds{1}_{B'})\big|_{f=0}\Gcal_{k-1}(\Gcal_\s(1|\cdot))\nonumber \\ 
&\quad+\delta \Gcal_\b(e^{-f}; \mathds{1}_B)\big|_{f=0}\delta (\Gcal_{k-1}(\Gcal_\s(e^{-f}|\cdot));\mathds{1}_{B'})\big|_{f=0}\nonumber \\
&\quad+\delta \Gcal_\b(e^{-f}; \mathds{1}_{B'})\big|_{f=0}\delta (\Gcal_{k-1}(\Gcal_\s(e^{-f}|\cdot));\mathds{1}_{B})\big|_{f=0} \nonumber\\
&\quad + \Gcal_\b(1)\delta^2 (\Gcal_{k-1}(\Gcal_\s(e^{-f}|\cdot)); \mathds{1}_B,\mathds{1}_{B'})\big|_{f=0},
\intertext{where the differentiation rule \eqref{eq:muderivative} yields}
&\mu^{(2)}_{k|k-1}(B\times B') = \mu^{(2)}_\b(B\times B')\nonumber
\\ 
&\quad-\mu_\b(B)\delta (\Gcal_{k-1}(\Gcal_\s(e^{-f}|\cdot));\mathds{1}_{B'})\big|_{f=0}\nonumber\\
&\quad-\mu_\b(B')\delta (\Gcal_{k-1}(\Gcal_\s(e^{-f}|\cdot));\mathds{1}_{B})\big|_{f=0}\nonumber\\
&\quad +  \delta^2 (\Gcal_{k-1}(\Gcal_\s(e^{-f}|\cdot)); \mathds{1}_B,\mathds{1}_{B'})|_{f=0}, \label{eq:secondordertotal_last} 
\end{align}
\end{subequations}
where $\mu_{\b}$ and $\mu^{(2)}_{\b}$ are the first- and second-order moment measures of the birth process, respectively. Let us first focus on the term $\delta (\Gcal_{k-1}(\Gcal_\s(e^{-f}|\cdot));\mathds{1}_{B})\big|_{f=0}$ in \eqref{eq:secondordertotal_last}. Using the definition of the \ac{pgfl} \eqref{eq:pgfl} we can write
\begin{subequations} \label{eq:firstorderterm}
  \begin{align}
    &\delta (\Gcal_{k-1}(\Gcal_\s(e^{-f}|\cdot));\mathds{1}_{B})\big|_{f=0} \nonumber
    \\
    &= \sum_{n\geq 0} \int_{\Xcal^n} \hspace*{-5pt} \delta \left.\left(\left[\prod_{i=1}^n \Gcal_\s(e^{-f}|x_i) \right] ;\mathds{1}_{B}\right)\right|_{f=0} \hspace*{-15pt} P^{(n)}_{k-1}(\d x_{1:n})
    \\
    &\stackrel{\eqref{eq:productrule}}{=} \sum_{n\geq 0} \int_{\Xcal^n} \sum_{i=1}^n \left.\delta(\Gcal_\s(e^{-f}|x_i) ;\mathds{1}_{B})\right|_{f=0} P^{(n)}_{k-1}(\d x_{1:n})
    \\
    &\stackrel{\eqref{eq:campbell}}{=} \int \delta(\Gcal_\s(e^{-f}|x) ;\mathds{1}_{B})\big|_{f=0} \mu_{k-1}(\d x). \label{eq:firstorderterm_last}
  \end{align}
\end{subequations}
The survival process for a target with state $x$ at the previous time step can be described with a Bernoulli point process with parameter $p_{\s}(x)$ and spatial distribution $t(\cdot|x)$, and thus \eqref{eq:bernoulli} gives
\begin{equation}
  \Gcal_\s(e^{-f}|x) = 1 - p_{\s}(x) + p_{\s}(x)\int e^{-f(y)}t(\d y|x).
\end{equation}
It follows that
\begin{subequations} \label{eq:survivalprocessdecomp}
  \begin{align}
    \delta(\Gcal_\s(e^{-f}|x) ;\mathds{1}_{B}) &= p_{\s}(x)\int\delta(e^{-f(y)};\mathds{1}_{B}) t(\d y|x)
    \\
    &= - p_{\s}(x)\int\hspace{-0.1cm}\mathds{1}_{B}(y) e^{-f(y)} t(\d y|x),
  \end{align}
\end{subequations}
which leads to
\begin{equation} \label{eq:survivalprocessdecompatzero}
  \delta(\Gcal_\s(e^{-f}|x) ;\mathds{1}_{B})\big|_{f=0} = - p_{\s}(x)t(B|x).
\end{equation}
Substituting \eqref{eq:survivalprocessdecompatzero} in \eqref{eq:firstorderterm_last} yields
\begin{equation}
  \delta (\Gcal_{k-1}(\Gcal_\s(e^{-f}|\cdot));\mathds{1}_{B})\big|_{f=0} = -\hspace{-0.1cm}\int p_{\s}(x)t(B|x)\mu_{k-1}(\d x). \label{eq:firstorderterm_result}
\end{equation}
Let us write the last term $\delta^2 (\Gcal_{k-1}(\Gcal_\s(e^{-f}|\cdot)); \mathds{1}_B,\mathds{1}_{B'})|_{f=0}$ in \eqref{eq:secondordertotal_last} in a similar manner as above. From the definition of the \ac{pgfl} \eqref{eq:pgfl} we can write
\begin{subequations} \label{eq:secondorderterm}
  \begin{align}
    &\delta^2 (\Gcal_{k-1}(\Gcal_\s(e^{-f}|\cdot)); \mathds{1}_B,\mathds{1}_{B'})|_{f=0}
    \\
    &= \sum_{n\geq 0} \int_{\Xcal^n}\hspace{-0.1cm} \delta^2\hspace{-0.1cm} \left.\left(\left[\prod_{i=1}^n \Gcal_\s(e^{-f}|x_i) \right]; \mathds{1}_{B}, \mathds{1}_{B'}\right)\right|_{f=0} \hspace{-0.3cm}P^{(n)}_{k-1}(\d x_{1:n})
    \\
    &\stackrel{\eqref{eq:productrule}}{=} \sum_{n\geq 0} \int_{\Xcal^n} \sum_{i=1}^n \left.\delta^2(\Gcal_\s(e^{-f}|x_i) ;\mathds{1}_{B}, \mathds{1}_{B'})\right|_{f=0} P^{(n)}_{k-1}(\d x_{1:n}) \nonumber
    \\
    &+ \sum_{n\geq 0} \int_{\Xcal^n} \sum_{\substack{1 \leq i, j \leq n \\ i \neq j}} \delta(\Gcal_\s(e^{-f}|x_i) ;\mathds{1}_{B})\big|_{f=0} \nonumber
    \\
    &~\hspace{1.5cm}\cdot\delta(\Gcal_\s(e^{-f}|x_j); \mathds{1}_{B'})\big|_{f=0} P^{(n)}_{k-1}(\d x_{1:n})
    \\
    &\stackrel{\eqref{eq:campbell}}{=} \int \delta^2(\Gcal_\s(e^{-f}|x) ;\mathds{1}_{B}, \mathds{1}_{B'})\big|_{f=0} \mu_{k-1}(\d x) \nonumber
    \\
    &+ \int \delta(\Gcal_\s(e^{-f}|x) ;\mathds{1}_{B})\big|_{f=0} \nonumber
    \\
    &~\hspace{1.5cm}\cdot\delta(\Gcal_\s(e^{-f}|x') ;\mathds{1}_{B'})\big|_{f=0}  \nu^{(2)}_{k-1}(\d(x, x')). \label{eq:secondorderterm_last}
  \end{align}
\end{subequations}
From \eqref{eq:survivalprocessdecomp}, the value of $\delta^2(\Gcal_\s(e^{-f}|x) ;\mathds{1}_{B}, \mathds{1}_{B'})\big|_{f=0}$ is found to be
\begin{equation}
  \delta^2(\Gcal_\s(e^{-f}|x) ;\mathds{1}_{B}, \mathds{1}_{B'})\big|_{f=0} = p_{\s}(x)t(B \cap B'|x),
\end{equation}
so that \eqref{eq:secondorderterm_last} becomes
\begin{align} \label{eq:secondorderterm_result}
  &\delta^2 (\Gcal_{k-1}(\Gcal_\s(e^{-f}|\cdot)); \mathds{1}_B,\mathds{1}_{B'})|_{f=0}\nonumber
  \\
  &= \mu_{\mathrm{s}}(B \cap B')+\hspace{-0.1cm} \int p_{\s}(x)t(B|x)p_{\s}(x')t(B'|x') \nu^{(2)}_{k-1}(\d(x, x')).
\end{align}
Substituting \eqref{eq:firstorderterm_result} and \eqref{eq:secondorderterm_result} in \eqref{eq:secondordertotal_last} and setting $B = B'$ yields 
\begin{align}
&\mu^{(2)}_{k|k-1}(B\times B)\nonumber
\\ 
&\quad  = \mu^{(2)}_\b(B\times B)+2\mu_\b(B)\mu_{\mathrm{s}}(B)  + \mu_{\mathrm{s}}(B) \nonumber 
\\
&\quad +\int p_{\s}(x)t(B|x)p_{\s}(x')t(B|x') \nu^{(2)}_{k-1}(\d(x, x')). \label{eq:secondordertotal_result} 
\end{align}
Using the definition of the variance \eqref{eq:variance} then yields
\begin{align}
&\var_{k|k-1}(B) = \var_\b(B) + [\mu_\b(B)]^2 - [\mu_{k|k-1}(B)]^2\nonumber
\\ 
&\quad +2\mu_\b(B)\mu_\mathrm{s}(B) + \mu_\mathrm{s}(B) \nonumber 
\\
&\quad +\int p_{\s}(x)t(B|x)p_{\s}(x')t(B|x') \nu^{(2)}_{k-1}(\d(x, x')), \label{eq:variance_result} 
\end{align}
and substituting the expression of the predicted intensity \eqref{eq:phdpred} to $\mu_{k|k-1}(B)$ in \eqref{eq:variance_result} yields the desired result.
\end{proof}

\subsection{Proof of Cor.~\ref{cor:varpred}}
\begin{proof}
Let us assume that the probability of survival $p_{\s,k}$ is uniform over the state space.  First of all, Eq.~\eqref{eq:phdpredsurv} with $B = \Xcal$ simplifies to
\begin{subequations}
 \label{eq:phdpredsurv_simpl}
\begin{align}
\mu_{\mathrm{s},k}(\Xcal) &= p_{\s, k}\int \underbrace{t_{k|k-1}(\Xcal|x)}_{=1} \mu_{k-1}(\d x)
\\
&= p_{\s, k}\mu_{k-1}(\Xcal).
\end{align}
\end{subequations}
From Eq.~\eqref{eq:varpred} we can then compute the variance of the survival process $\var_{\s, k}$ in the whole state space, i.e.
\begin{subequations}
\begin{align}
&\var_{\s, k}(\Xcal) \nonumber
\\
&=  \mu_{\mathrm{s},k}(\Xcal) [1 - \mu_{\mathrm{s},k}(\Xcal)]
\\
 &+p_{\s, k}^2\int \underbrace{t_{k|k-1}(\Xcal|x)}_{=1}\underbrace{t_{k|k-1}(\Xcal|x')}_{=1} \nu_{k-1}^{(2)}(\d(x,x'))
 \\
&\stackrel{\eqref{eq:phdpredsurv_simpl}}{=} p_{\s, k}\mu_{k-1}(\Xcal) \Big[1 - p_{\s, k}\mu_{k-1}(\Xcal)\Big] + p_{\s, k}^2\nu_{k-1}^{(2)}(\Xcal \times \Xcal)
 \\
&\stackrel{\eqref{eq:decomp}}{=} p_{\s, k}\mu_{k-1}(\Xcal) \Big[1 - p_{\s, k}\mu_{k-1}(\Xcal)\Big]\nonumber
\\
&+ p_{\s, k}^2\left[\mu_{k-1}^{(2)}(\Xcal \times \Xcal) - \mu_{k-1}(\Xcal)\right]
\\
&\stackrel{\eqref{eq:variance}}{=}p_{\s, k}\mu_{k-1}(\Xcal) \Big[1 - p_{\s, k}\mu_{k-1}(\Xcal)\Big]\nonumber
\\
&+ p_{\s, k}^2\left[\var_{k-1}(\Xcal) + [\mu_{k-1}(\Xcal)]^2 - \mu_{k-1}(\Xcal)\right]
\\
&= p_{\s, k}^2\var_{k-1}(\Xcal) + p_{\s, k}[1 - p_{\s, k}] \mu_{k-1}(\Xcal).
\end{align}\qedhere
\end{subequations}

\end{proof}

\subsection{Proof of Thm.~\ref{thm:phdupdate}}
\begin{proof}
Let us denote by $\Gcal_{\c, k}$ the \ac{pgfl} of the clutter process, and by $\Gcal_{\d, k}$ the \ac{pgfl} of the process describing the detection (or not) of a target in scene. For the sake of simplicity, time subscripts on the predicted target process, clutter process, and detection process will be omitted in this proof. In particular, we shall use the short-hand notations $\alpha:=\alpha_{k|k-1}$, $\beta:=\beta_{k|k-1}$, $s := s_{k|k-1}$, and $\mu:=\mu_{k|k-1}$ for the quantities describing the predicted target process $\Phi_{k|k-1}$. In addition, we shall use the short-hand notation $q_\d(x) := 1 - p_{\d}(x)$ to denote the probability of missed detection for a target with state $x \in \Xcal$.

From Assumptions~\ref{assum:2} we can write the explicit formulation of the joint \ac{pgfl} describing the predicted target process and the observation collected from the sensor \cite{Mahler2003Multitarget}:
\begin{equation} \label{eq:jointpgfl}
  \Gcal_{\J,k}(g,h) = \Gcal_{k|k-1}\left(h \Gcal_\d(g|\cdot) \right) \Gcal_\c(g),
\end{equation}
where the multiplicative form stems from the independence between the target-generated measurements and the clutter measurements; the composition appears because the detection process applies to each target described by the predicted target process $\Phi_{k|k-1}$. Since both the predicted target process and the clutter process are assumed Panjer, \eqref{eq:jointpgfl} takes the more specific form
\begin{equation} \label{eq:jointpgfl2}
  \Gcal_{\J,k}(g,h) = \mu(\Xcal)^{\alpha}\Big(F_\d(g,h)\Big)^{-\alpha}\Big(F_\c(g)\Big)^{-\alpha_\c},
\end{equation}
where
\begin{subequations} \label{eq:F1}
 \begin{align}
  \hspace*{-10pt}F_\d(g,h) &:= \mu(\Xcal)\left(1 + \frac{1}{\beta}\int(1 - h(x)\Gcal_\d(g|x))s(\d x)\right)
  \\
  &= \int \left[1 + \frac{1 - h(x)\Gcal_\d(g|x)}{\beta}\right]\mu(\d x),
\end{align}
\end{subequations}
and
\begin{equation} \label{eq:F2}
  F_\c(g) := 1 + \frac{1}{\beta_\c}\int(1 - g(z))s_\c(z)\d z.
\end{equation}
Note that the expression of the clutter term \eqref{eq:F2} follows directly from the definition of a Panjer process \eqref{eq:panjer}; the detection term \eqref{eq:jointpgfl2} stems from \eqref{eq:panjer} as well but is then scaled by the predicted mean number of targets $\mu(\Xcal)$, so that the final result of the theorem exploits similar notations as the \ac{cphd} filter in \cite{Vo2007Analytic}, \cite{Delande2014Regional}. The detection process for a target with state $x$ can be described with a Bernoulli point process with parameter $p_{\d}(x)$ and spatial distribution density $l(\cdot|x)$, and thus \eqref{eq:bernoulli} gives
\begin{equation}
\Gcal_\d(g|x) = q_\d(x)+ p_\d(x) \int_\Zcal g(z)l(z|x)\d z.
\end{equation}
Note that both $F_\d$ and $F_\c$ are linear \ac{wrt} to the argument $g$, and thus only their first-order derivatives are non-zero; given an arbitrary measurement $z \in Z_k$, we can write
\begin{align}
  \delta F_\d(g,h; \delta_z) &= -\int \frac{h(x)p_\d(x){l(z|x)}}{\beta} \mu(\d x), \label{eq:F3}
  \\
  \delta F_\c(g;\delta_z) &= -\frac{1}{\beta_\c}s_\c(z). \label{eq:delF2}
\end{align}
Similarly to the \ac{phd} filter update \cite{Mahler2003Multitarget}, the \ac{pgfl} of the updated target process $\Phi_k$ is obtained from the differentiation of the joint \ac{pgfl} \eqref{eq:jointpgfl2} using Bayes' rule:
\begin{equation}
\label{eq:condupdatemodel}
\Gcal_{k}(h) = \cfrac{\delta^{|Z_k|}\Gcal_{\J,k} (g,h; (\delta_{z})_{z\in Z_k})|_{g=0}}{\delta^{|Z_k|}\Gcal_{\J,k} (g,1; (\delta_{z})_{z\in Z_k}))|_{g=0}}.
\end{equation}
Using the higher-order product \eqref{eq:generalproductrule} and chain \eqref{eq:generalchainrule} rules, the $|Z_k|$-th derivative of the joint \ac{pgfl} \eqref{eq:jointpgfl2} in directions $(\delta_{z})_{z\in Z_k}$ yields
\begin{subequations}
\label{eq:mthorder}
\begin{align}
&\delta^{|Z_k|}\Gcal_{\J,k}(g,h; (\delta_{z})_{z\in Z_k}) \nonumber
\\
&= \mu(\Xcal)^{\alpha} \sum_{j=0}^{|Z_k|} \frac{(\alpha)_j}{\beta^j}\frac{(\alpha_\c)_{|Z_k|-j}}{\beta_\c^{|Z_k|-j}}F_\d(g,h)^{-\alpha-j}F_\c(g)^{-\alpha_\c-|Z_k|+j} \nonumber
\\
& \qquad \cdot \sum_{\substack{Z\subseteq Z_k \\ |Z| = j}} \left( \prod_{z\in Z} F_\d^z(h) \prod_{z' \in Z_k \backslash Z} s_\c(z') \right) \label{eq:mthorder1}
\\
& \propto \sum_{j=0}^{|Z_k|} \frac{(\alpha)_j}{\beta^j}\frac{(\alpha_\c)_{|Z_k|-j}}{(\beta_\c F_\c(g))^{|Z_k|-j}}F_\d(g,h)^{-j} \sum_{\substack{Z\subseteq Z_k \\ |Z| = j}}\prod_{z\in Z} \frac{F_\d^z(h)}{s_{\c}(z)}, \label{eq:mthorder2}
\end{align}
\end{subequations}
where
\begin{align}
F_\d^z(h) :=  \int h(x)p_\d(x){l(z|x)}\mu(\d x).
\end{align}
The proportional constant in \eqref{eq:mthorder} is the quantity $\mu(\Xcal)^{\alpha} F_\d(g,h)^{-\alpha}F_\c(g)^{-\alpha_\c}\prod_{z\in Z_k}s_{\c}(z)$; since it is discarded in the ratio \eqref{eq:condupdatemodel}, it will be omitted from now on. Details of the developments leading to \eqref{eq:mthorder} can be found in Lem.~VI.6 in \cite{Schlangen2016PHD}, where a similar result is produced. 

Similarly to \cite{Mahler2003Multitarget}, we can finally compute the intensity of the updated target process $\Phi_k$ in any region $B \in \Bcal(\Xcal)$ from the first-order derivative of its \ac{pgfl} \eqref{eq:condupdatemodel}, i.e. 
\begin{equation} \label{eq:updated_intensity}
  \mu_k(B) = \cfrac{\delta^{|Z_k| + 1}\Gcal_{\J,k} (g,h; (\delta_{z})_{z\in Z_k}, \mathds{1}_B)|_{g=0, h=1}}{\delta^{|Z_k|}\Gcal_{\J,k} (g,1; (\delta_{z})_{z\in Z_k}))|_{g=0}}.
\end{equation}
We first need to differentiate \eqref{eq:mthorder} in direction $\mathds{1}_B$ through the product rule \eqref{eq:productrule} and get
\begin{equation}
\label{eq:mplusone}
\begin{split}
&\delta^{|Z_k|+1}\Gcal_{\J,k}(g,h; (\delta_{z})_{z\in Z_k},\mathds{1}_B) \propto (-\beta \delta F_\d(g,h; \mathds{1}_B))  \\
&\cdot \sum_{j=0}^{|Z_k|} \frac{(\alpha)_{j+1}}{\beta^{j+1}}\frac{(\alpha_\c)_{|Z_k|-j}}{(\beta_\c F_\c(g))^{|Z_k|-j}}F_\d(g,h)^{-j-1}\sum_{\substack{Z\subseteq Z_k \\ |Z| = j}} \prod_{z\in Z}F_\d^z(h) 
\\&+\!\sum_{z\in Z_k}F_\d^z(\mathds{1}_B)\!\! \sum_{j=0}^{|Z_k|-1}\frac{(\alpha)_{j+1}}{\beta^{j+1}}\frac{(\alpha_\c)_{(|Z_k|-1)-j}}{(\beta_\c F_\c(g))^{(|Z_k|-1)-j}}  \\
&\cdot F_\d(g,h)^{-j-1} \sum_{\substack{Z\subseteq Z_k\backslash \{z\}\\|Z|=j}}\prod_{z' \in Z_k}F_\d^{z'}(h),
\end{split}
\end{equation}
where
\begin{equation}
\begin{split}
& \delta F_\d(g,h; \mathds{1}_B)\\
&= -\frac{1}{\beta}\int_B \bigg[q_\d(x)+p_\d(x) \int_\Zcal g(z)\ell(z|x) \d z \bigg] \mu(\d x).
\end{split}
\end{equation}
Substituting \eqref{eq:mthorder} and \eqref{eq:mplusone} into \eqref{eq:updated_intensity} yields the desired result.
\end{proof}
\subsection{Proof of Thm.~\ref{thm:varupdate}}
\begin{proof}
The variance $\var_k$ of the updated target process $\Phi_k$ in an arbitrary region $B \in \Bcal(\Xcal)$ can be computed from the first- and second-order moment measures $\mu_k, \mu^{(2)}_k$ through the decomposition \eqref{eq:variance}. We have already computed the first-order moment measure $\mu_k$ in Thm.~\ref{thm:phdupdate}, and we shall now focus on the expression of the second-order moment measure $\mu^{(2)}_k$.

From \eqref{eq:muderivative}, we can compute the second-order moment measure $\mu^{(2)}_k(B \times B')$ in any regions $B, B' \in \Bcal(\Xcal)$ from the second-order derivative of the Laplace functional $\Lcal_k$ of the updated target process $\Phi_k$. Substituting $\exp(-f)$ to $h$ in the  \ac{pgfl} \eqref{eq:condupdatemodel} yields the expression of the Laplace functional $\Lcal_k$, and from \eqref{eq:muderivative} it follows that \cite{Delande2014Regional}
\begin{equation} \label{eq:updated_2nd_measure}
  \mu^{(2)}_k(B \times B') = \cfrac{\delta^{|Z_k| + 2}\Gcal_{\J,k} (0, e^{-f}; (\delta_{z})_{z\in Z_k}, \mathds{1}_B, \mathds{1}_{B'})|_{g=0, f=0}}{\delta^{|Z_k|}\Gcal_{\J,k} (g, 1; (\delta_{z})_{z\in Z_k}))|_{g=0}}.
\end{equation}
The denominator in \eqref{eq:updated_2nd_measure} has already been computed in \eqref{eq:mthorder}; we shall thus focus here on the derivation in directions $\mathds{1}_B, \mathds{1}_{B'}$ of the numerator
\begin{align}
&\delta^{|Z_k|}\Gcal_{\J,k}(0,e^{-f}; (\delta_{z})_{z\in Z_k}) \nonumber
\\
&\propto \sum_{j=0}^{|Z_k|} \frac{(\alpha)_j}{\beta^j}\frac{(\alpha_\c)_{|Z_k|-j}}{(1 + \beta_\c)^{|Z_k|-j}}F_\d(0,e^{-f})^{-j}\sum_{\substack{Z\subseteq Z_k \\ |Z| = j}}\prod_{z\in Z} \frac{F_\d^z(e^{-f})}{s_{\c}(z)}. \label{eq:mthorderLaplace}
\end{align}
The first-order derivative of \eqref{eq:mthorderLaplace} in direction $\mathds{1}_B$ is
\begin{equation}
\label{eq:firstderiv}
\begin{split}
&\delta^{|Z_k|+1}\Gcal_{\J,k}(0,e^{-f}; (\delta_{z})_{z\in Z_k},\mathds{1}_B) \\
&\propto -\sum_{j=0}^{|Z_k|}  \frac{(\alpha)_{j+1}}{\beta^{j+1}}\frac{(\alpha_\c)_{|Z_k|-j}}{(\beta_\c+1)^{|Z_k|-j}}F_\d(0,e^{-f})^{-j-1}\\
&\qquad \cdot F_\mathrm{md}(e^{-f} \mathds{1}_B)\sum_{\substack{Z\subseteq Z_k \\ |Z| = j}} \prod_{z\in Z}F_\d^z(e^{-f})\\
&\quad-\sum_{j=1}^{|Z_k|}\frac{(\alpha)_{j}}{\beta^{j}}\frac{(\alpha_\c)_{|Z_k|-j}}{(\beta_\c+1)^{|Z_k|-j}}F_\d(0,e^{-f})^{-j}\\
&\quad ~\cdot\sum_{z\in Z_k}\frac{F_\d^z(e^{-f} \mathds{1}_B)}{s_\c(z)}\sum_{Z\subseteq Z_k\backslash \{z\}}\prod_{z' \in Z} \frac{F_\d^{z'}(e^{-f})}{s_\c(z)},
\end{split}
\end{equation}
where
\begin{equation}
F_\mathrm{md}(h) := \int h(x)q_\d(x)\mu(\d x).
\end{equation}
The second-order derivative of \eqref{eq:mthorderLaplace} in directions $\mathds{1}_B,\mathds{1}_{B'}$ then takes the form \eqref{eq:secondderiv}.

\setcounter{equation}{88}
Note that the third and fifth terms in \eqref{eq:secondderiv} correspond exactly to the updated first-order moment of the process. 
Subsituting and \eqref{eq:mthorder} and  \eqref{eq:secondderiv} into \eqref{eq:updated_2nd_measure} yields
\begin{equation}
\label{eq:secondordermoment}
\begin{split}
\mu_k^{(2)}(B \times B') =&~\mu_k(B\cap B')+\mu_k^\phi(B)\mu_k^\phi(B')\ell_2(\phi)\\
&+\mu_k^\phi(B) \sum_{z\in Z}\frac{\mu_k^z(B')}{s_\c(z)}\ell_{2}(z)\\
&+\mu_k^\phi(B') \sum_{z\in Z}\frac{\mu_k^z(B)}{s_\c(z)}\ell_{2}(z)\\
&+\sum_{z,z'\in Z_k}~\frac{\mu_k^z(B)}{s_\c(z)}\frac{s_k^{z'}(B')}{s_\c(z')} \ell_{2}^{\neq}(z,z').
\end{split}
\end{equation}
Following \eqref{eq:variance}, the intensity \eqref{eq:phdupdate} is then squared and subtracted from the second-order moment \eqref{eq:secondordermoment} evaluated with $B' = B$ in order to yield the desired quantity $\var_k(B)$.
\end{proof}

\subsection{Proof of Cor.~\ref{cor:limit}}
\begin{proof}
Let us assume that the predicted target process $\Phi_{k|k-1}$ is Poisson with rate $\lambda_{k|k-1}$, i.e., $\alpha_{k|k-1}, \beta_{k|k-1} \rightarrow \infty$, with constant ratio $\lambda_{k|k-1} = \frac{\alpha}{\beta}$. For the same of simplicity, the time subscripts on $\alpha_{k|k-1}, \beta_{k|k-1}, \lambda_{k|k-1}$ are omitted for the rest of the proof. Note first that, since $\mu(\d x) = \lambda s(\d x)$, we have
\begin{subequations}
  \begin{align}
    \lim_{\alpha, \beta \rightarrow \infty} F_{\d} &= \lim_{\alpha, \beta \rightarrow \infty} \int \bigg[1 + \underbrace{\frac{p_{\d, k}(x)}{\beta}}_{\rightarrow 0}\bigg] \lambda s(\d{x})
    \\
    &= \lambda.
  \end{align}
\end{subequations}

In order to check the convergence of the intensity update equation \eqref{eq:phdupdate}, we only need to check the convergence of the term \eqref{eq:ypsilon} as it is the only term that contains $\alpha$ or $\beta$. We can write:
\begin{subequations}
\label{eq:ypsilon_limit}
\begin{align}
&\lim_{\alpha, \beta \rightarrow \infty} \Upsilon_{u}(Z) \nonumber
\\
&= \lim_{\alpha, \beta \rightarrow \infty} \sum_{j=0}^{|Z|} \frac{(\alpha)_{j+u}}{(\beta)^{j+u}} \frac{(\alpha_\c)_{|Z|-j}}{(\beta_\c+1)^{|Z|-j}} F_\d^{-j-u}~e_{j}(Z)\\
&\stackrel{\eqref{eq:pochhammer}}{=} \lim_{\alpha, \beta \rightarrow \infty} \sum_{j=0}^{|Z_k|}\lambda \bigg(\lambda + \underbrace{\frac{1}{\beta}}_{\rightarrow 0}\bigg)\dots \bigg(\lambda + \underbrace{\frac{j+u-1}{\beta}}_{\rightarrow 0}\bigg)\nonumber\\
&\quad\cdot \frac{(\alpha_\c)_{|Z|-j}}{(\beta_\c+1)^{|Z|-j}}\underbrace{F_{\d}^{-j - u}}_{\rightarrow \lambda^{-j - u}}e_{j}(Z)\\
&= \sum_{j=0}^{|Z|}\frac{(\alpha_\c)_{|Z|-j}}{(\beta_\c+1)^{|Z|-j}}e_{j}(Z).
\end{align}
\end{subequations}
Note in particular that the limit of $\Upsilon_{u}(Z)$ is independent of the value of $u$; the corrective terms \eqref{eq:corrective1} thus converge to
\begin{equation}
\left\{
\begin{aligned}
\lim_{\alpha, \beta \rightarrow \infty} \ell_1(\phi) &= 1
\\
\lim_{\alpha, \beta \rightarrow \infty} \ell_1(z) &= \frac{\sum_{j=0}^{|Z_k| - 1}\frac{(\alpha_\c)_{|Z_k|-j-1}}{(\beta_\c+1)^{|Z_k|-j - 1}}e_{j}(Z_k \setminus \{z\})}{\sum_{j=0}^{|Z_k|}\frac{(\alpha_\c)_{|Z_k|-j}}{(\beta_\c+1)^{|Z|-j}}e_{j}(Z_k)},
\end{aligned}
\right.
\end{equation}
which coincides with the results of Thm~III.2 in \cite{Schlangen2016PHD}.

If we further assume that the clutter process is Poisson, the intensity update equation \eqref{eq:phdupdate} further converges to the intensity update equation of the original \ac{phd} filter, as shown in \cite{Schlangen2016PHD}.
\end{proof}

\subsection{Proof of Prop.~\ref{prop:covariance}}
\begin{proof}
The covariance is found with Eq.~\eqref{eq:covariance}. For the second-order \ac{phd} filter, the first- and second-order moment measures are given by Eqns \eqref{eq:phdupdate} and \eqref{eq:secondordermoment}. For the \ac{phd} filter, they are given by (28) and (31) in \cite{Delande2014Regional}, and for the \ac{cphd} filter by (19) and (29) ibid.
\end{proof}


 \section*{Appendix B: Second-order \ac{gm}-\ac{phd} filter} \label{AppAlgorithms}
        
    \begin{algorithm}[H]
      \begin{algorithmic}
	\State \emph{Input}
	  \State Collection of terms: $\left\{{\mu_k^z(\Xcal)}\right\}_{z \in Z_k}$
	\\
	\State \emph{Vieta's theorem}
	\State Expand: $p(x) = \prod_{z\in Z_k}\left(x-{\mu_k^z(\Xcal)}\right) = \sum_{j=0}^{m_k} p_j x^j$
	\State Set $e_j(Z_k) = p_j$ for all $0 \leq j \leq m_k$
	\\
	\State \emph{Output}
	\State $\{e_j(Z_k)\}_{0\leq j \leq m_k}$ 
      \end{algorithmic}
      \caption{Function to compute the elementary symmetric functions using Vieta's theorem.\label{alg:vieta}}
    \end{algorithm}
 
    \begin{algorithm}[H]
      \begin{algorithmic}
	\State \emph{Input}
	  \State Posterior: $\{w_{k-1}^{(i)},m_{k-1}^{(i)},P_{k-1}^{(i)}\}_{i=1}^{N_{k-1}}, \var_{k-1}(\Xcal)$
	  \State Births: $\{w_{\b,k-1}^{(i)},m_{\b,k-1}^{(i)},P_{\b,k-1}^{(i)}\}_{i=1}^{N_{\b,k-1}}, \var_{\b,k}(\Xcal)$
	\\
	\State \emph{Survival process}
	\State $\mu_{k-1}(\Xcal) = \sum_{i=1}^{N_{k-1}}w_{k-1}^{(i)}$
	\For{$1 \leq i \leq n_{k-1}$}
	    \State $w_{k|k-1}^{(i)} = p_{\s,k} w_{k-1}^{(i)}$
	    \State $m_{k|k-1}^{(i)} = F_{k-1} m_{k-1}^{(i)}$
	    \State $P_{k|k-1}^{(i)} = F_{k-1} P_{k-1}^{(i)} F_{k-1}^T+Q_{k-1}$
	\EndFor
	\State $\var_{\s, k}(\Xcal) = p_{\s,k}^2\var_{k-1}(\Xcal) + p_{\s,k}[1- p_{\s, k}]\mu_{k-1}(\Xcal)$
	\\
	\State \emph{Newborn process}
	\For{$1 \leq j \leq N_{\b,k-1}$}
	    \State $w_{k|k-1}^{(n_{k-1}+j)} = w_{\b,k-1}^{(j)}$
	    \State $m_{k|k-1}^{(n_{k-1}+j)} = m_{\b,k-1}^{(j)},\quad P_{k|k-1}^{(n_{k-1}+j)} = P_{\b,k-1}^{(j)}$
	\EndFor
	\\
	\State $N_{k|k-1} = N_{k-1}+N_{\b,k-1}$	
	\State $\var_{k|k-1}(\Xcal) = \var_{\b, k}(\Xcal) + \var_{\s, k}(\Xcal)$
	\\
	\State \emph{Output}
	\State Prediction: $\{w_{k|k-1}^{(i)},m_{k|k-1}^{(i)},P_{k|k-1}^{(i)}\}_{i=1}^{N_{k|k-1}},  \var_{k|k-1}(\Xcal)$ 
      \end{algorithmic}
      \caption{Time prediction (time $k$).}
    \end{algorithm}

    \begin{algorithm}[H]
      \begin{algorithmic}
	\State \emph{Input}
	  \State Prediction: $\{w_{k|k-1}^{(i)},m_{k|k-1}^{(i)},P_{k|k-1}^{(i)}\}_{i=1}^{N_{k|k-1}},  \var_{k|k-1}(\Xcal)$ 
	  \State Current measurements: $Z_k = \{ z_j\}_{j=1}^{M_k}$
	\\
	\State \emph{Panjer parameters}
	\State $\mu_{k|k-1}(\Xcal) = \sum_{i=1}^{N_{k|k-1}} w_{k|k-1}^{(i)}$
	\State $\alpha_{k|k-1} = \mu_{k|k-1}(\Xcal)^2/(\var_{k|k-1}(\Xcal)-\mu_{k|k-1}(\Xcal))$
	\State $\beta_{k|k-1} = \mu_{k|k-1}(\Xcal)/(\var_{k|k-1}(\Xcal)-\mu_{k|k-1}(\Xcal))$
	\\
	\State \emph{Missed detection and measurement terms}
	\For{$1 \leq i \leq N_{k|k-1}$}
		\State $w_{\phi,k}^{(i)} = (1-p_{\d,k}) w_{k|k-1}^{(i)} $
		\State $m_{\phi,k}^{(i)} = m_{k|k-1}^{(i)} , \quad P_{\phi,k}^{(i)} = P_{k|k-1}^{(i)} $
	\EndFor
	\State $\mu_k^\phi(\Xcal) = (1-p_{\d,k})\mu_{k|k-1}(\Xcal)$
	\For{$1 \leq j \leq M_k$}
	  	\For{$1 \leq i \leq N_{k|k-1}$}
	    		\State $y_{k}^{(i,j)} = z_j - H_k m_{k|k-1}^{(i)} $ 
	    		\State $S_{k}^{(i)} = H_k P_{k|k-1}^{(i)} H_k^{T} + R_k$ 
	    		\State $K_{k}^{(i)} = P_{k|k-1}^{(i)} H_k^{T} [S_{k}^{(i)}]^{-1}$
	    		\State $w_{\d,k}^{(i,j)} = p_{\d,k}w_{\d,k|k-1}^{(i,j)}\Ncal(z;y_{k}^{(i,j)},S_{k}^{(i)})/s_{\c,k}$
	    		\State $m_{\d,k}^{(i,j)} = m_{k|k-1}^{(i)} + K_{k}^{(i)}y_{k}^{(i,j)}$
	    		\State $P_{\d,k}^{(i,j)} = (I-K_{k}^{(i)}H_k)P_{k|k-1}^{(i)}$
		\EndFor
		\State $\mu_k^{z_j}(\Xcal) = \sum_{i=1}^{N_{k|k-1}} w_{\d,k}^{(i,j)}$
	\EndFor
	\algstore{bkbreak}
       \end{algorithmic}
     \end{algorithm}
     \addtocounter{algorithm}{-1}
     \begin{algorithm}[H]
       \begin{algorithmic}
 	\algrestore{bkbreak}
	\State \emph{Corrective terms}
	\State $F_\d = (1+\frac{p_{\d,k}}{\beta_{k|k-1}})\sum_{z\in Z_k}\mu_k^{z_j}(\Xcal)$
	\State Compute $\{e_d(Z_k)\}_{0\leq d \leq M_k}$ using Alg.~\ref{alg:vieta}
	\For{$0 \leq u \leq 2$}
	\State $\Upsilon_{u}(Z_k) = \sum_{j=0}^{M_k} \frac{(\alpha_{k|k-1})_{j+u}}{(\beta_{k|k-1})^{j+u}} \frac{(\alpha_{\c, k})_{m_k-j}}{(\beta_{\c, k} + 1)^{m_k-j}}F_\d^{-j-u}e_{j}(Z_k)$ 
	\EndFor
	  \State $\ell_1(\phi) := \Upsilon_{1}(Z_k)/\Upsilon_{0}(Z_k), \quad \ell_2(\phi) := \Upsilon_{2}(Z_k)/\Upsilon_{0}(Z_k)$ 
	  \For{$1 \leq i \leq M_k$}
	  	\State Compute $\{e_d(Z_k\setminus z_i)\}_{0\leq d \leq M_k-1}$ using Alg.~\ref{alg:vieta}
	    \For{$1 \leq u \leq 2$}
			\State $\Upsilon_{u}(Z_k\setminus z_i) = \sum_{d=0}^{M_k-1} \frac{(\alpha_{k|k-1})_{d+u}}{(\beta_{k|k-1})^{d+u}} $
			\State \hfill $\cdot\frac{(\alpha_{\c, k})_{m_k-1-d}}{(\beta_{\c, k} + 1)^{m_k-1-d}}F_\d^{-d-u}e_{d}(Z_k\setminus z_i)$ 
		\EndFor
	    	 \State $\ell_1(z_i) := \Upsilon_{1}(Z_k\setminus z_i)/\Upsilon_{0}(Z_k)$
	    	 \State $\ell_2(z_i) := \Upsilon_{2}(Z_k\setminus z_i)/\Upsilon_{0}(Z_k)$
	 	\For{$1\leq i < j \leq M_k$}
	 	\State Compute $\{e_d(Z_k\setminus\{z_i,z_j\})\}_{0\leq d \leq M_k-2}$ using Alg.~\ref{alg:vieta}
		  \State $\Upsilon_{2}(Z_k\setminus \{z_i,z_j\}) = \sum_{d=0}^{M_k-2} \frac{(\alpha_{k|k-1})_{d+2}}{(\beta_{k|k-1})^{d+2}} $
		  \State \hfill $\cdot\frac{(\alpha_{\c, k})_{m_k-2-d}}{(\beta_{\c, k} + 1)^{m_k-2-d}}F_\d^{-d-2}e_{d}(Z_k\setminus \{z_i,z_j\})$
		 \State $\ell^{\neq}_2(z_i,z_j) = \Upsilon_{2}(Z_k\setminus \{z_i,z_j\})/\Upsilon_{0}(Z_k)$ 
	      \EndFor
	  \EndFor
	  \\
	\State \emph{Missed detection terms}
	\For{$1\leq i \leq N_{k|k-1}$}
	  	\State $w_{k}^{(i)} = \ell_1(\phi)w_{\phi,k}^{(i)}$
	  	\State $m_k^{(i)} = m_{\phi,k}^{(i)}$
	  	\State $P_k^{(i)} = P_{\phi,k}^{(i)}$
	  	\State \emph{Association terms}
	  	\For{$1\leq j \leq M_{k}$}
	  		\State $w_{k}^{(i\cdot n_{k|k-1}+j)} = \ell_1(z_j)w_{\d,k}^{(i,j)}$
	  		\State $m_{k}^{(i\cdot n_{k|k-1}+j)} = m_{\d,k}^{(i,j)}$
	  		\State $P_{k}^{(i\cdot n_{k|k-1}+j)} = P_{\d,k}^{(i,j)}$	  					\EndFor
	\EndFor
	\State $N_k = N_{k|k-1} + N_{k|k-1} M_k$
	\State $\mu_k(\Xcal) = \sum_{i=1}^{N_k} w_k^{(i)}$
	\\
	\State \emph{Variance update}
	  \State \vspace{-0.4cm}\[\begin{split}
&\var_k(\Xcal)=\mu_k(\Xcal)
+\mu_k^\phi(\Xcal)^2\left[\ell_2(\phi)-\ell_1(\phi)^2\right] \\
&+2\mu_k^\phi(\Xcal)\sum_{z\in Z_k}~\frac{\mu_k^z(\Xcal)}{s_{\c,k}(z)}\left[\ell_{2}(z)-\ell_1(\phi)\ell_{1}(z)\right]\\
&+\sum_{z\neq z'\in Z_k}~\frac{\mu_k^z(\Xcal)}{s_{\c,k}(z)}\frac{\mu_k^{z'}(\Xcal)}{s_{\c,k}(z')}\left[\ell_{2}^{\neq}(z,z')-\ell_{1}(z)\ell_{1}(z')\right],
\end{split}\]
	\\
	\State \emph{Output}
		  \State Posterior: $\{w_{k}^{(i)},m_{k}^{(i)},P_{k}^{(i)}\}_{i=1}^{N_{k}}, \var_{k}(\Xcal)$
      \end{algorithmic}
      \caption{Data update (time $k$).}
    \end{algorithm}

\newpage
\bibliographystyle{IEEEtran}
\bibliography{refs}

{\onecolumn

\setcounter{equation}{87}
 \begin{figure*}
\hspace{-0.5cm}\noindent\rule{0.9\paperwidth}{0.4pt}
\begin{equation}
\label{eq:secondderiv}
 \begin{split}
&\delta^{|Z_k|+2}\Gcal_{\J,k}(0,e^{-f}; (\delta_{z})_{z\in Z_k},\mathds{1}_B,\mathds{1}_{B'}) \\
 &\propto~ \sum_{j=0}^{|Z_k|} \frac{(\alpha)_{j+2}}{\beta^{j+2}}\frac{(\alpha_\c)_{|Z_k|-j}}{(\beta_\c+1)^{|Z_k|-j}}F_\d(0,e^{-f})^{-j-2} F_\mathrm{md}(e^{-f}\mathds{1}_B)F_\mathrm{md}(e^{-f}\mathds{1}_{B'})\sum_{\substack{Z\subseteq Z_k \\ |Z| = j}} \prod_{z\in Z}\frac{F_\d^z(e^{-f})}{s_\c(z)}\\
 &\quad+ \sum_{j=1}^{|Z_k|} \frac{(\alpha)_{j+1}}{\beta^{j+1}}\frac{(\alpha_\c)_{|Z_k|-j}}{(\beta_\c+1)^{|Z_k|-j}}F_\d(0,e^{-f})^{j-1} F_\mathrm{md}(e^{-f}\mathds{1}_B) \sum_{z\in Z_k}\frac{F_\d^z(e^{-f}\mathds{1}_{B'})}{s_\c(z)}\sum_{\substack{Z\subseteq Z_k\backslash \{z\}\\|Z|=j-1}}\prod_{z' \in Z} \frac{F_\d^{z'}(e^{-f})}{s_\c(z)}\\
 &\quad+\sum_{j=0}^{|Z_k|} \frac{(\alpha)_{j+1}}{\beta^{j+1}}\frac{(\alpha_\c)_{|Z_k|-j}}{(\beta_\c+1)^{|Z_k|-j}}F_\d(0,e^{-f})^{j-1} F_\mathrm{md}(e^{-f}\mathds{1}_{B\cap B'})
 \sum_{\substack{Z\subseteq Z_k \\ |Z| = j}} \prod_{z\in Z}\frac{F_\d^z(e^{-f})}{s_\c(z)}\\
 &\quad+\sum_{j=1}^{|Z_k|}  \frac{(\alpha)_{j+1}}{\beta^{j+1}}\frac{(\alpha_\c)_{|Z_k|-j}}{(\beta_\c+1)^{|Z_k|-j}}F_\d(0,e^{-f})^{j-1} F_\mathrm{md}(e^{-f}\mathds{1}_{B'})\sum_{z\in Z_m}\frac{F_\d^z(e^{-f}\mathds{1}_B)}{s_\c(z)}\sum_{\substack{Z\subseteq Z_k\backslash \{z\}\\|Z|=j-1}}\prod_{z' \in Z} \frac{F_\d^{z'}(e^{-f})}{s_\c(z)}\\
 &\quad+\sum_{j=1}^{|Z_k|} \frac{(\alpha)_{j}}{\beta^{j}}\frac{(\alpha_\c)_{|Z_k|-j}}{(\beta_\c+1)^{|Z_k|-j}}F_\d(0,e^{-f})^{j}\sum_{z\in Z_k}\frac{F_\d^z(e^{-f}\mathds{1}_{B \cap B'})}{s_\c(z)}
 \sum_{\substack{Z\subseteq Z_k\backslash \{z\}\\|Z|=j-1}}\prod_{z' \in Z} \frac{F_\d^{z'}(e^{-f})}{s_\c(z)}\\
 &\quad+\sum_{j=2}^{|Z_k|} \frac{(\alpha)_{j}}{\beta^{j}}\frac{(\alpha_\c)_{|Z_k|-j}}{(\beta_\c+1)^{|Z_k|-j}}F_\d(0,e^{-f})^{j} \sum_{\substack{z,z'\in Z_k\\z\neq z'}}\frac{F_\d^z(e^{-f}\mathds{1}_B)}{s_\c(z)} 
 \frac{F_\d^{z'}(e^{-f}\mathds{1}_{B'})}{s_\c(z)}
 \sum_{\substack{Z\subseteq Z_k\backslash \{z,z'\}\\|Z|=j-2}}\prod_{z'' \in Z} \frac{F_\d^{z''}(e^{-f})}{s_\c(z)}.
\end{split}
\end{equation}
\hspace{-0.5cm}\noindent\rule{0.9\paperwidth}{0.4pt}
\end{figure*}


\begin{figure}
\centering
\begin{subfigure}[t]{0.45\linewidth}
\centering
\input{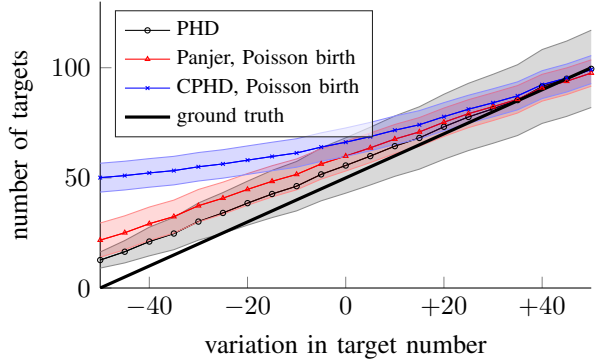}
\caption{Experiment 1.1.}
\label{fig:pois_mean}
\end{subfigure}~~
\begin{subfigure}[t]{0.45\linewidth}
\centering
\input{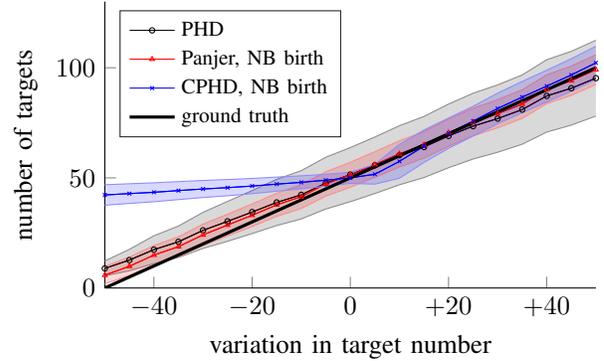}
\caption{Experiment 1.2.}
\label{fig:nb_mean}
\end{subfigure}\\
\begin{subfigure}[t]{0.45\linewidth}
\centering
\input{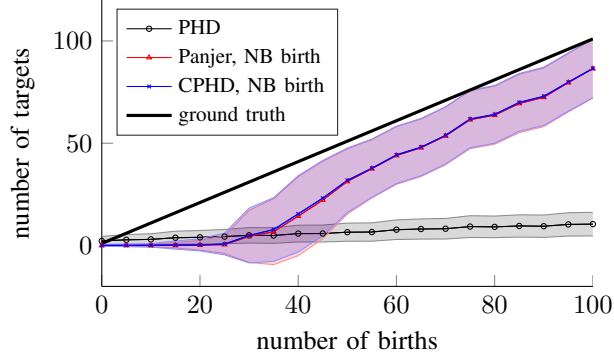}
\caption{Experiment 1.3.}
\label{fig:bigbirth_mean}
\end{subfigure}
\caption{Results for Scenario 1, averaged over $20$ \ac{mc} runs. The lines depict the mean of the estimated number of targets, the coloured areas show the $2\sigma$ confidence region (estimated by the filter).}
\label{fig:birth_death_scenario}
\end{figure}

\begin{figure}
\centering
\begin{subfigure}[t]{0.45\linewidth}
\centering
\input{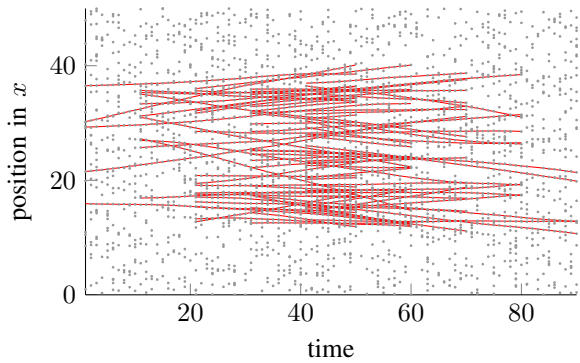}
\caption{Position in $x$ over time.}
\label{fig:step_setup_x}
\end{subfigure}~~
\begin{subfigure}[t]{0.45\linewidth}
\centering
\input{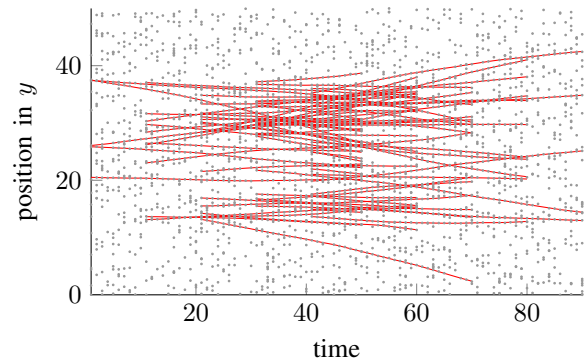}
\caption{Position in $y$ over time.}
\label{fig:step_setup_y}
\end{subfigure}
\caption{The setup of experiment 2.1, plotted separately for $x$ and $y$ over time (shown for one \ac{mc} run). The ground truth is plotted in red, the measurements in grey.}
\label{fig:step_gt}
\end{figure}

\begin{figure}
\centering
\begin{subfigure}[t]{0.45\linewidth}
\centering
\input{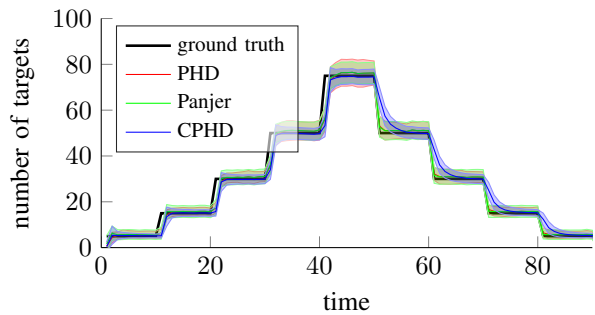}
\caption{Estimated target number, experiment 2.1.}
\label{fig:NB_card}
\end{subfigure}~~~
\begin{subfigure}[t]{0.45\linewidth}
\centering
\input{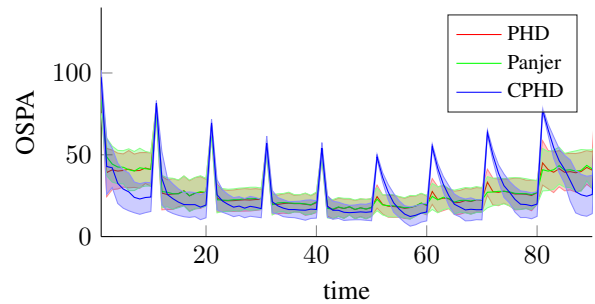}
\caption{OSPA results, experiment 2.1.}
\label{fig:NB_ospa}
\end{subfigure}\\
\begin{subfigure}[t]{0.45\linewidth}
\centering
\input{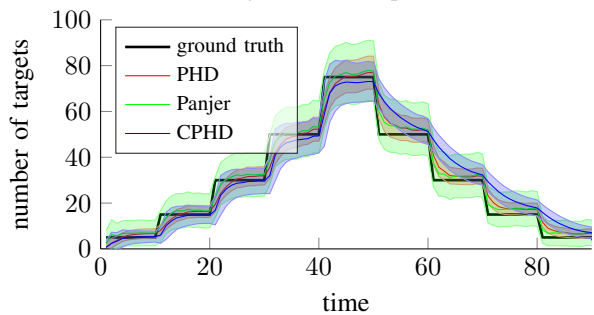}
\caption{Estimated target number, experiment 2.2.}
\label{fig:pd_card}
\end{subfigure}~~~
\begin{subfigure}[t]{0.45\linewidth}
\centering
\input{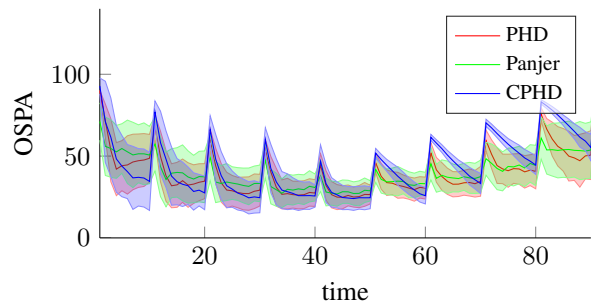}
\caption{OSPA results, experiment 2.2.}
\label{fig:pd_ospa}
\end{subfigure}
\caption{Results for Scenario 2, averaged over $100$ \ac{mc} runs.  Fig.~\ref{fig:NB_card} and \ref{fig:pd_card} show the estimated means and variances of the number of targets, Fig.~\ref{fig:NB_ospa} and \ref{fig:pd_ospa} displays the mean and standard deviation of the respective OSPA results. The rows depict the results of experiments 2.1 ($p_{\d} = 0.95$) and 2.2 ($p_{\d} = 0.6$), respectively.}
\label{fig:step}
\end{figure}

\begin{figure}
\centering
\begin{subfigure}[t]{0.45\linewidth}
\centering
\input{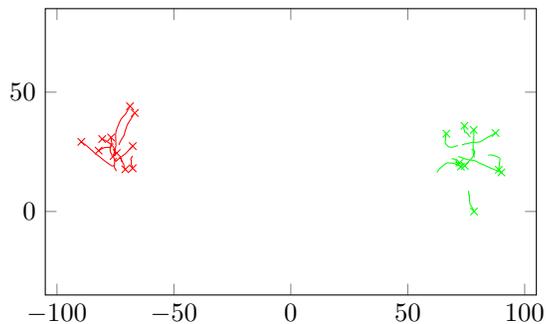}
\caption{Tracking scenario, with region $A$ on the left and region $B$ on the right.}
\label{fig:scenario}
\end{subfigure}\\
\begin{subfigure}[t]{0.45\linewidth}
\centering
%
%
\begin{tikzpicture}

\begin{axis}[%
width=0.8\linewidth,
height=1.5in,
at={(2.6in,1.146979in)},
scale only axis,
xmin=0,
xmax=100,
xlabel={time},
ymin=-0.8,
ymax=0.8,
ylabel={$\corr(A,B)$},
axis x line*=bottom,
axis y line*=left,
legend style={at={(0.97,0.97)},anchor=north east,legend cell align=left,align=left,draw=white!15!black,fill opacity=0.6,draw opacity=1,text opacity=1,font=\footnotesize}
]
\addplot [color=red,solid]
  table[row sep=crcr]{%
1	0\\
2	0\\
3	0\\
4	0\\
5	0\\
6	0\\
7	0\\
8	0\\
9	0\\
10	0\\
11	0\\
12	0\\
13	0\\
14	0\\
15	0\\
16	0\\
17	0\\
18	0\\
19	0\\
20	0\\
21	0\\
22	0\\
23	0\\
24	0\\
25	0\\
26	0\\
27	0\\
28	0\\
29	0\\
30	0\\
31	0\\
32	0\\
33	0\\
34	0\\
35	0\\
36	0\\
37	0\\
38	0\\
39	0\\
40	0\\
41	0\\
42	0\\
43	0\\
44	0\\
45	0\\
46	0\\
47	0\\
48	0\\
49	0\\
50	0\\
51	0\\
52	0\\
53	0\\
54	0\\
55	0\\
56	0\\
57	0\\
58	0\\
59	0\\
60	0\\
61	0\\
62	0\\
63	0\\
64	0\\
65	0\\
66	0\\
67	0\\
68	0\\
69	0\\
70	0\\
71	0\\
72	0\\
73	0\\
74	0\\
75	0\\
76	0\\
77	0\\
78	0\\
79	0\\
80	0\\
81	0\\
82	0\\
83	0\\
84	0\\
85	0\\
86	0\\
87	0\\
88	0\\
89	0\\
90	0\\
91	0\\
92	0\\
93	0\\
94	0\\
95	0\\
96	0\\
97	0\\
98	0\\
99	0\\
100	0\\
};
\addlegendentry{PHD};

\addplot [color=green,solid]
  table[row sep=crcr]{%
1	0.352254449545914\\
2	0.127412009698277\\
3	0.0671729395093314\\
4	0.0589288486206651\\
5	0.0560375035941623\\
6	0.0550436965639697\\
7	0.0553692306367644\\
8	0.0545275285794901\\
9	0.0551389711145034\\
10	0.056050806994374\\
11	0.0922721506390587\\
12	0.0594588456672515\\
13	0.0567272301645794\\
14	0.0563190575243907\\
15	0.0554356456095122\\
16	0.0559402932879057\\
17	0.0562273787679425\\
18	0.0552504678131484\\
19	0.055687152258647\\
20	0.057508139249773\\
21	0.0896820085865593\\
22	0.0588215733200545\\
23	0.0554468866197616\\
24	0.0555025317200749\\
25	0.0567193894318584\\
26	0.0560434486907051\\
27	0.0560306911349522\\
28	0.0561447693085918\\
29	0.0560550288792442\\
30	0.0565158059545182\\
31	0.0897043071127013\\
32	0.0584220481533465\\
33	0.0556680817821633\\
34	0.0562731256180023\\
35	0.0554469722792738\\
36	0.0546817813634624\\
37	0.0553543468659792\\
38	0.0555990676746921\\
39	0.054546256288028\\
40	0.0564668139088953\\
41	0.0884599024172379\\
42	0.0596956065940409\\
43	0.056175383379609\\
44	0.0550778202431228\\
45	0.0541930974307768\\
46	0.0551072822793982\\
47	0.0547911638240976\\
48	0.054693242183283\\
49	0.0555979158333539\\
50	0.0564907030925241\\
51	0.0880887891924541\\
52	0.0582742280088652\\
53	0.0552120545717628\\
54	0.0554812613962561\\
55	0.0548608094086335\\
56	0.0538077850818397\\
57	0.0547494204793057\\
58	0.0549794023771243\\
59	0.0542564391887247\\
60	0.0555476819248698\\
61	0.0878574225707657\\
62	0.0582994072788905\\
63	0.0550778291180293\\
64	0.0559898399974227\\
65	0.0553390002125133\\
66	0.0568397820134914\\
67	0.0560102393215455\\
68	0.0557214588677339\\
69	0.0551550960758234\\
70	0.0559690305684464\\
71	0.0883508524394891\\
72	0.0597975009151036\\
73	0.0555726804144036\\
74	0.0569214436018857\\
75	0.0560323201664469\\
76	0.0554887313087434\\
77	0.0563357450830563\\
78	0.0557718332040293\\
79	0.0566983840231045\\
80	0.0547352168271487\\
81	0.085801745050488\\
82	0.0557195043141414\\
83	0.0531176950166322\\
84	0.0521667663140372\\
85	0.0523464746897163\\
86	0.0530122314654709\\
87	0.0531591824739879\\
88	0.0524233247512562\\
89	0.0527359823871808\\
90	0.0545883941664938\\
91	0.0883044128227975\\
92	0.0562502439097635\\
93	0.0533972707246117\\
94	0.0523438536718761\\
95	0.0515860053248415\\
96	0.0512956311019127\\
97	0.0505497431500615\\
98	0.0519881395486387\\
99	0.0509026864155229\\
100	0.0521896268228381\\
};
\addlegendentry{Panjer};

\addplot [color=blue,solid]
  table[row sep=crcr]{%
1	0.348446998123648\\
2	0.175585526009128\\
3	-0.097623992300727\\
4	-0.230225653985263\\
5	-0.316670383062984\\
6	-0.356762885889439\\
7	-0.387230373640109\\
8	-0.379946724941325\\
9	-0.390304485479039\\
10	-0.633564264355875\\
11	-0.16124819351488\\
12	-0.239399793264263\\
13	-0.249845722232672\\
14	-0.277448468288152\\
15	-0.29438683116451\\
16	-0.338088220604589\\
17	-0.313925748540143\\
18	-0.359171423054503\\
19	-0.369484016748995\\
20	-0.636683669717306\\
21	-0.111719235127583\\
22	-0.20402511728375\\
23	-0.236753543189486\\
24	-0.29974024706147\\
25	-0.342977324485984\\
26	-0.336338035549546\\
27	-0.342111836338902\\
28	-0.363492798131204\\
29	-0.391875040909066\\
30	-0.634242269346517\\
31	-0.0942235994579206\\
32	-0.21728637385533\\
33	-0.261273754795042\\
34	-0.252224164554701\\
35	-0.278308537517438\\
36	-0.351433924553638\\
37	-0.355694170221123\\
38	-0.31948556487658\\
39	-0.341856439705582\\
40	-0.637837823022444\\
41	-0.20057241038042\\
42	-0.238233195308472\\
43	-0.2250368753384\\
44	-0.196498510994098\\
45	-0.315297325640658\\
46	-0.32887547781466\\
47	-0.350313924138631\\
48	-0.360855629380146\\
49	-0.370965978142604\\
50	-0.637596218007115\\
51	-0.0657069234110783\\
52	-0.236071139226696\\
53	-0.248531752410723\\
54	-0.266175517455729\\
55	-0.2867785216546\\
56	-0.35381096896121\\
57	-0.361690468686343\\
58	-0.348687006247842\\
59	-0.374200059799725\\
60	-0.64315491880092\\
61	-0.0815187271023569\\
62	-0.203515198361334\\
63	-0.258931110730568\\
64	-0.32385923002408\\
65	-0.351174604034938\\
66	-0.313228634971792\\
67	-0.342236092022622\\
68	-0.347921671003428\\
69	-0.34740882913678\\
70	-0.636502182488698\\
71	-0.203441567069744\\
72	-0.221155121796281\\
73	-0.236897827125603\\
74	-0.23426099782076\\
75	-0.291609065363475\\
76	-0.329833149075289\\
77	-0.376354343788339\\
78	-0.358639552604233\\
79	-0.346948446550466\\
80	-0.593419303377878\\
81	-0.0900022819959485\\
82	-0.182901211870694\\
83	-0.166908942638137\\
84	-0.265764758044087\\
85	-0.333961781924739\\
86	-0.372547405779331\\
87	-0.333511127392446\\
88	-0.3670790127508\\
89	-0.354992626773961\\
90	-0.597058347048836\\
91	-0.0636488131275441\\
92	-0.193199373641757\\
93	-0.228108978700573\\
94	-0.235033849762752\\
95	-0.291394413762813\\
96	-0.272802156257505\\
97	-0.333279883945411\\
98	-0.343718923126658\\
99	-0.325338841489223\\
100	-0.553522547906867\\
};
\addlegendentry{CPHD};

\end{axis}
\end{tikzpicture}%
\caption{Correlation between the estimated number of targets in regions $A$ and $B$.}
\label{fig:correlation}
\end{subfigure}~~
\begin{subfigure}[t]{0.45\linewidth}
\centering
\input{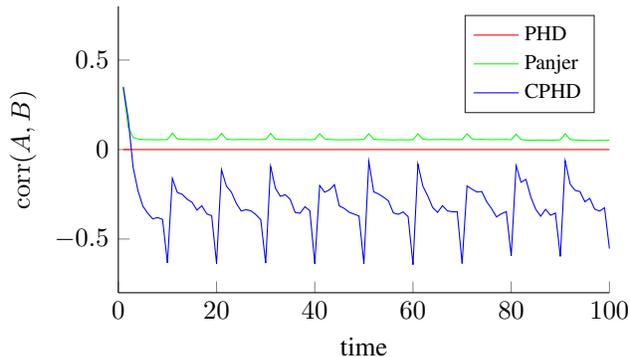}
\caption{Mean and standard deviation of the estimated target number, \ac{phd} filter.}
\label{fig:card_phd}
\end{subfigure}\\
\begin{subfigure}[t]{0.45\linewidth}
\centering
\input{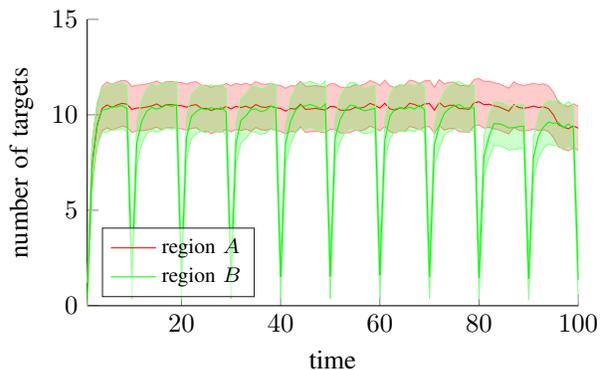}
\caption{Mean and standard deviation of the estimated target number, Panjer filter.}
\label{fig:card_panjer}
\end{subfigure}~~
\begin{subfigure}[t]{0.45\linewidth}
\centering
\input{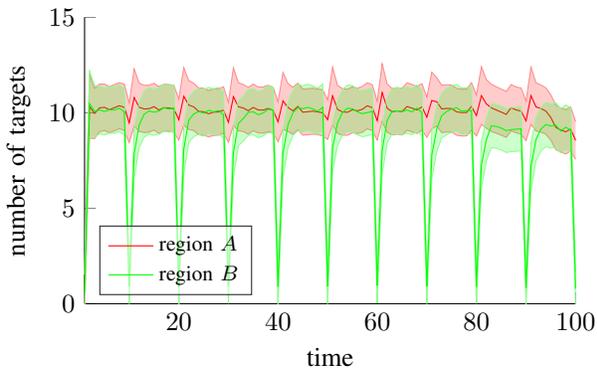}
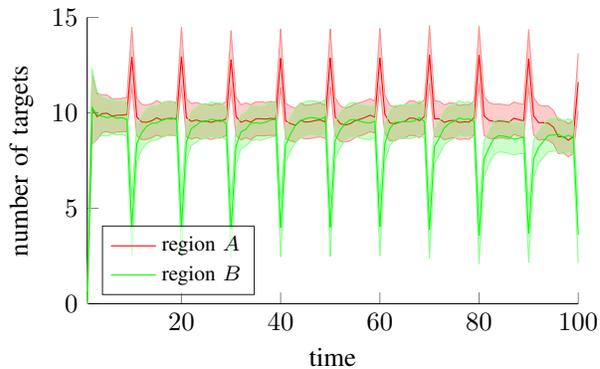
\caption{Mean and standard deviation of the estimated target number, \ac{cphd} filter.}
\label{fig:card_cphd}
\end{subfigure}
\caption{Results for Scenario 3, averaged over $100$ \ac{mc} runs. 
Fig.~\ref{fig:correlation} shows the correlation  in $A$ and $B$ for all filters. Fig.~\ref{fig:card_phd}, \ref{fig:card_panjer} and \ref{fig:card_cphd} depict the mean and standard deviation of the estimated number of targets per region for the three filters.}
\label{fig:spooky}
\end{figure}
\twocolumn
}

\end{document}